\documentclass[11pt]{article}

\usepackage{epstopdf,float}
\usepackage{multirow}
\usepackage{graphics}
\usepackage{amsmath,amsthm,amsfonts}
\usepackage{subfigure}
\usepackage{color, setspace, multirow}
\usepackage[T1]{fontenc}
\usepackage[utf8]{inputenc}
\usepackage{authblk} 
\usepackage{natbib} 
\usepackage{graphicx}
\usepackage{array}
\usepackage{mwe,hyperref}
\usepackage{graphbox, lineno}
\usepackage[export]{adjustbox}
\usepackage[normalem]{ulem}
 \usepackage{enumitem,rotating}
\usepackage{caption}

\newcommand{\bfD}{{\bf D}}

\newcommand{\bfX}{{\bf X}}

\newcommand{\bftheta}{\mbox{\boldmath $\theta$}}
\newcommand{\bfepsilon}{\mbox{\boldmath $\epsilon$}}

\newcommand{\bfmu}{\mbox{\boldmath $\mu$}}
\newcommand{\bfomega}{\mbox{\boldmath $\omega$}}
\newcommand{\bfphi}{\mbox{\boldmath $\phi$}}

\newcommand{\bfbeta}{\mbox{\boldmath $\beta$}}

\newcommand{\bfalpha}{\mbox{\boldmath $\alpha$}}

\newcommand{\bfgamma}{\mbox{\boldmath $\gamma$}}

\newcommand{\bfZ}{{\bf Z}}
\newcommand{\bfY}{{\bf Y}}

\newcommand{\bfA}{{\bf A}}

\newcommand{\bfs}{{\bf s}}

\newcommand\redsout{\bgroup\markoverwith{\textcolor{red}{\rule[0.5ex]{2pt}{0.4pt}}}\ULon}

\newcommand{\blind}{1}

\def\spacingset#1{\renewcommand{\baselinestretch}%
	{#1}\small\normalsize} \spacingset{1}

\begin{document}

\if1\blind
{
	\title{Two-stage MCMC for Fast Bayesian Inference of Large Spatio-temporal Ordinal Data, with Application to US Drought}
	\author{Staci Hepler, Rob Erhardt$^{\dagger}$\\
		Department of Statistical Sciences, Wake Forest University, U.S.
	}
	\maketitle
} \fi

\if0\blind
{
	\title{\bf A }
	\maketitle
} \fi

\bigskip
\begin{abstract}
High dimensional space-time data pose known computational challenges when fitting spatio-temporal models.  Such data show dependence across several dimensions of space as well as in time, and can easily involve hundreds of thousands of observations.  Many spatio-temporal models result in a dependence structure across all observations and can be fit only at a substantial computational cost, arising from dense matrix inversion, high dimensional parameter spaces, poor mixing in Markov Chain Monte Carlo, or the impossibility of utilizing parallel computing due to a lack of independence anywhere in the model fitting process. These computational challenges are exacerbated when the response variable is ordinal, and especially as the number of ordered categories grows.  Some spatio-temporal models achieve computational feasibility for large datasets but only through overly restrictive model simplifications, which we seek to avoid here.  In this paper we demonstrate a two-stage algorithm to fit a Bayesian spatio-temporal model to large datasets when the response variable is ordinal.  
The first stage models locations independently in space, capturing temporal dependence, and can be run in parallel.  The second stage resamples from the first stage posterior distributions with an acceptance probability computed to impose spatial dependence from the full spatio-temporal model.  The result is fast Bayesian inference which samples from the full spatio-temporal posterior and is computationally feasible even for large datasets.  We quantify the substantial computational gains our approach achieves, and demonstrate the preservation of the posterior distribution as compared to the more costly single-stage model fit.  We apply our approach to a large spatio-temporal drought dataset in the United States, a dataset too large for many existing spatio-temporal methods.

\end{abstract}

\noindent%
\vfill

$\dagger$ Corresponding author.  Address 1834 Wake Forest Road, Winston-Salem NC, 27109, US.  Email \texttt{erhardrj@wfu.edu}

\newpage
\spacingset{1.45} 

\section{Introduction}

Spatio-temporal data arise in a wide array of applications and scientific domains including, but not limited to, environmental science, ecology, epidemiology, and neuroscience.  
The analysis of spatio-temporal data requires statistical models that account for spatial and temporal dependencies in the physical process of interest. There are many choices and consequences for the assumed structure of spatio-temporal models. In the spatial setting, Gaussian process models are often employed with geostatistical data collected over continuous spatial domains, whereas Markov random field models are commonly assumed for areal data on a spatial lattice. There are various approaches to extend these to the spatio-temporal setting when data are indexed by both location and time. For example, Gaussian processes can be defined over a 3-dimensional domain such that time is the third dimension, with corresponding covariance functions specified to account for dependence across both space and time. When time is discrete, a common modeling strategy is to assume an autoregressive or Markov dependence structure in time. 

Regardless of the assumed structure, modeling spatio-temporal data presents several computational challenges. The sheer volume of data can be immense, requiring substantial storage capabilities. Additionally, the spatial and temporal dependencies are typically captured through large, dense covariance matrices, requiring considerable computational resources for inference in both frequentist and Bayesian approaches. These computational challenges can be further exacerbated for non-Gaussian data. Numerous approaches have been suggested in the literature to alleviate the computational expense required for statistical modeling of spatial and spatio-temporal data. In what follows, we provide a brief and non-exhaustive overview of relevant literature, and we refer the reader to \cite{heaton2019case} for an overview and comparison of various spatial models. 


Reduced rank models alleviate the computational burden by approximating the spatial process with a lower-dimension representation. This is often achieved through the use of spatial basis functions  \citep{billings2002smooth, nychka2000spatial, cressie2008fixed}, and has been extended to the spatio-temporal setting \citep{cressie2010fixed, katzfuss2012bayesian, zammit2021frk}. Predictive process models reduce the dimension by projecting the spatial or spatio-temporal process onto a lower-dimensional subspace \citep{banerjee2008gaussian, finley2012bayesian}. Other approaches have replaced dense covariance or precision matrices with sparse approximations \citep{furrer2006covariance, kaufman2008covariance, rue2005gaussian}. Vecchia approximations induce sparsity by assuming the joint distribution is a product of conditional distributions, where each conditional distribution depends only on a small subset of observations \citep{katzfuss2021general, pan2025block}. \cite{peruzzi2022highly} extend this notion and propose meshed Gaussian processes, which assume a sparsity inducing directed acyclic graph on a partition of the spatial domain. Nearest neighbor Gaussian process models can be used as sparsity inducing priors for a high dimensional spatial or spatio-temporal models \citep{datta2016hierarchical, datta2016nonseparable}.

The methods proposed in this paper are applicable to Bayesian (non-Gaussian) dynamical spatio-temporal models (DSTM) on a spatial lattice with a first-order temporal autoregressive structure presented as outlined in \cite[Section 7.2.2]{cressie2015statistics}. That is, let $\mathcal{D}_s = \{\bfs_1,...,\bfs_n\}$ denote the spatial lattice, and let $D_t = \mathbb{N}$ denote the discrete temporal domain, indexed by the positive integers. For $\bfs \in \mathcal{D}_s$ and $t \in \mathcal{D}_t$, let $Y(\bfs, t)$ denote the observed (non-Gaussian) data. We assume $Y(\bfs, t) \mid Z(\bfs, t) \sim f\left(\cdot \mid Z(\bfs, t)\right)$ for some suitable $f(\cdot)$, where $Z(\bfs, t)$ is the spatio-temporal process. For example, in the case of a binary $Y(\bfs, t)$, one may assume $Y(\bfs, t) \mid Z(\bfs, t) = \mathbb{I}(Z(\bfs, t)>0)$ where $\mathbb{I}(\cdot)$ is the indicator function. When the observed response is a count variable, one may model $Y(\bfs, t) \mid Z(\bfs, t) \sim \text{Poisson}\left(\exp(Z(\bfs, t)) \right)$. This paper will consider an application of ordinal spatio-temporal data with $J+1$ ordered categories, and we will assume $Y(\bfs, t) = \sum_{j=0}^J j \cdot \mathbb{I}\left(\alpha_j < Z(\bfs, t) \leq \alpha_{j+1}\right)$. Define $\bfZ_t = \left( Z(\bfs_1, t),...,Z(\bfs_n, t)\right)'$ and $\bfmu_t = \left( \mu(\bfs_1, t),...,\mu(\bfs_n,t) \right)'$. The latent spatio-temporal process is assumed to be a DSTM of the form 
\begin{equation}
    \bfZ_t - \bfmu_t = \mathbf{M}\left(\bfZ_{t-1} - \bfmu_{t-1}  \right) + \bfepsilon_t,
    \label{eq:DSTM}
\end{equation}
where $\mathbf{M} = \text{diag}\left(\rho_1,...,\rho_n \right)$ and $\bfepsilon_t$ is a random vector for the errors. The trend is specified such that $\mu(\bfs, t) = \bfX(\bfs, t)' \bfbeta\left( \bfs\right)$, where $\bfX(\bfs, t)'$ is a $p$-dimensional vector of covariates at location $\bfs$ during time $t$, and the intercept and slope parameters in $\bfbeta\left( \bfs \right)$ are spatially varying and modeled with spatial random effects. 


Recursive Markov chain Monte Carlo (MCMC) algorithms have recently been developed to overcome computational challenges in some complex Bayesian models. \cite{lunn2013fully} originally developed the two-stage MCMC approach for Bayesian meta-analyses. The idea is to fit a simplified and computationally efficient model in stage one, which is often obtained by assuming independence, and then to use those stage one posterior samples as proposed values in a Metropolis-Hastings update of the full model in stage two. The assumption of independence permits the utilization of parallel computing to efficiently sample from the stage one posterior distributions. \cite{hooten2019making} refers to this approach as \textit{proposal-recursive Bayesian inference}. \cite{hooten2016hierarchical} illustrate how the two-stage approach can be applied to Bayesian hierarchical point process models of telemetry data, where the stage one models assume all individuals are independent. In this paper, we show how the proposal recursive MCMC approach permits computationally efficient inference for Bayesian DSTM models as in equation \eqref{eq:DSTM}, eliminating the need for approximations or dimension reduction. 

Our work is motivated by Bayesian spatio-temporal modeling of ordinal data, where the ordinal response indicates levels of drought severity and is recorded on a spatial lattice at discrete time periods. Our problem is high-dimensional in space and time, and a statistical model needs to account for spatial and temporal autocorrelation. The computational complexity precludes traditional Bayesian inference, and fitting the Bayesian hierarchical model in a traditional, ``single-stage'' MCMC algorithm is computationally prohibitive. This problem is exacerbated by the fact that we have ordinal data, which has unique computational challenges (see, e.g. \cite{schliep2015data}). 

The spatio-temporal ordinal drought data that motivates our analysis is introduced and described in Section \ref{se:data}. In Section \ref{se:model}, we describe the full spatio-temporal model that we assume and outline the two-stage MCMC procedure we will use to fit the model. Section \ref{se:comparison} contains a comparison of our two-stage MCMC procedure to a standard single-stage MCMC procedure for a subset of the data. We apply our model and proposed two-stage MCMC procedure to the full ordinal drought dataset in Section \ref{se:application}, and we illustrate its utility for forecasting drought in future time periods.

\section{Ordinal Drought Data}
\label{se:data} 
Here we briefly describe the data which motivates the methodology.  The data were compiled for this project and are freely available at \url{https://datadryad.org/stash/dataset/doi:10.5061/dryad.g1jwstqw7}, with a full description of the original sources, pre-processing steps, and exploratory data analysis available in \cite{erhardt2024homogenized}.  The response variable is an ordinal measure of drought called the US Drought Monitor (USDM).  The US Drought Monitor is jointly produced by the National Drought Mitigation Center at the University of Nebraska-Lincoln, the United States Department of Agriculture, and the National Oceanic and Atmospheric Administration.  This data product measures the severity of drought in the United States, with levels ranging from 0 (no drought) to D0 (pre-drought) to levels D1 through D4, corresponding to increasing severity of drought.  The USDM has been updated weekly since January 2000.  Although the USDM is a continuous measure defined everywhere in the US, we first discretized the spatial support over the contiguous United States into grid cells of size 0.5 degrees latitude by 0.5 degrees, longitude using a process described in full detail in \cite{erhardt2024homogenized}.  The result is an ordinal measure of drought $Y_{i,t}$ with six possible levels, and in this paper we model this for $i = 1, ..., I=3254$ areal spatial units across $t=1, ..., T=600$ weeks.  We will use the first 587 weeks for training data, and the final 13 weeks for test data in a forecasting application.

\begin{figure}[!h]
{\includegraphics[width=6.5in]{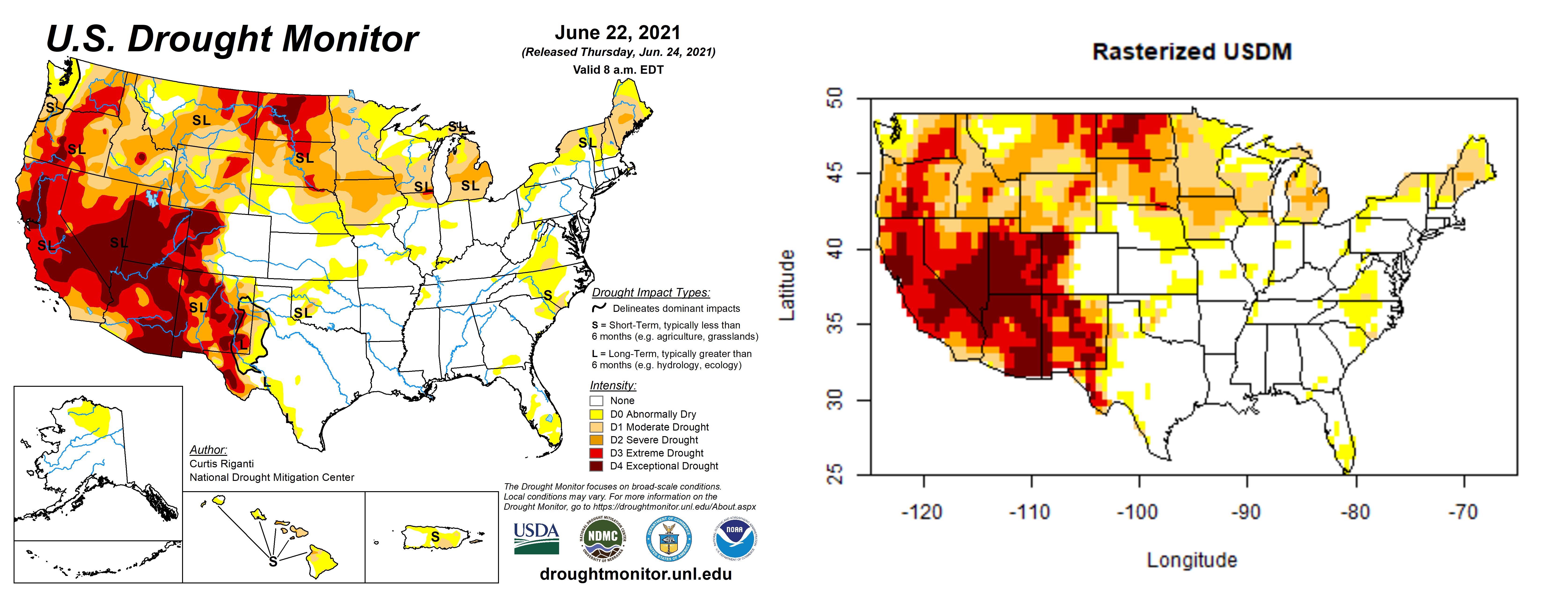}}
{\caption{Left panel: Raw data for the U.S. Drought Monitor for June 22, 2021.  
Map courtesy of NDMC.  Right panel: Discretized data from the same time period on a 0.5 degree spatial support. 
 This is one realization of the ordinal response variable over the $t=1, ..., 600$ total time periods.  The explanatory variables have the same spatial and temporal support.}\label{fig:usdm1}}
\end{figure}

Next we merged in three covariates related to drought and drought prediction.  These were obtained from the North America Land Data Assimilation System Phase 2 (NLDAS-2), an integrated observation and model reanalysis data set designed to drive offline land surface models \citep{mitchell2004multi, xia2012continental}.  Specifically, we used: weekly average evapotranspiration (\texttt{evp}, [kg/m2]), weekly average soil moisture content (\texttt{soilm}, [kg/m2]), and weekly average soil temperature (\texttt{tsoil}, [K]).  These were chosen to provide a mixture of atmospheric and land variables with varying levels of predictability, and also to ensure pairs that are both positively correlated (\texttt{evp} and \texttt{soilm}; \texttt{evp} and \texttt{tsoil}) and negatively correlated (\texttt{soilm} and \texttt{tsoil}), which cover important considerations for our prediction methodology described in section 5.2.  We upscaled each NLDAS-2 variable from its finer, native spatio-temporal resolution to the coarser 0.5 degree, weekly resolution to match the USDM data.  We call the vector of covariates $\bfX_{i,t}$.  Thus, all data in this application $Y_{i,t}$ and $\bfX_{i,t}$ have the common spatio-temporal weekly, 0.5 degree by 0.5 degree support for $i$ and $t$.

To recap, our dataset contains an ordinal response variable along with  continuous covariates \texttt{evp, soilm, tsoil}, for $I=3254$ spatial areal units and $T=600$ time periods (of which 587 are used as training data, and 13 are used as test data), for a combined 1,952,400 observational units of location-week.  These data display spatial dependence with nearby locations showing the strongest dependence, which weakens as the distance between the locations increases.  These data also display temporal autocorrelation, where both the response variable drought as well as the explanatory variables display varying levels of autocorrelation in time.  In Section 4 we demonstrate that a standard, \textit{single-stage} Markov chain Monte Carlo (MCMC) algorithm is not computationally feasible for this Bayesian spatio-temporal model, but that our two-stage MCMC algorithm is both computationally feasible and able to produce a spatio-temporal model fit capturing all data dependencies.

\section{Bayesian Modeling and Computation}
\label{se:model}

\subsection{Bayesian Spatio-temporal Model for Ordinal Data}

Let $Y_{i,t} \in \{0,1,...,J\}$ be the response which takes one of $J+1$ ordered levels, for location $i=1, ..., I$ and time period $t=1, ..., T$.  We assume there is a latent, continuous process, $Z_{i,t}$, such that $$Y_{i,t} = \sum_{j=0}^J j \cdot \textrm{I}(\alpha_j < Z_{i,t} \leq \alpha_{j+1}),$$ 
where $\bfalpha = (-\infty, \alpha_1, \alpha_2, ..., \alpha_J, \infty)$. 
The latent continuous drought process $Z_{i,t}$ is modeled with spatio-temporal structure and also depends on external covariates. More specifically, we assume
\begin{equation} 
Z_{i,t} = \begin{cases} 
 \bfX_{i,t} \bfbeta_{i} + \epsilon_{i,t} & \text{ for } t=1 \\
 \bfX_{i,t} \bfbeta_{i} + \rho_i \left(Z_{i, t-1} -\bfX_{i,t-1} \bfbeta_{i} \right) + \epsilon_{i,t}
 & \text{ for } t > 1,
\end{cases}
\label{eq:Zmod}
\end{equation}
where $\bfX_{i,t}$ is a $(P+1)$-dimensional vector of covariates including a 1 for an intercept parameter, $\bfbeta_{i} = (\beta_{0i}, \beta_{1i}, ..., \beta_{Pi})'$ is a location-specific $(P+1)$-dimensional parameter vector where each regression coefficient is a spatial random effect, $\rho_i$ is a spatially-varying temporal autoregressive term, and the errors are normally distributed and independent as $\epsilon_{i,t} \overset{ind}{\sim}\mathcal{N}(0, \sigma^2_i)$ with location-specific variances $\sigma^2_i$.  

To make this model identifiable, some constraints are needed on the parameters.  As one example why, consider that one could add a constant $h$ to each of the $\alpha_j$ as well as to the intercept of $\beta_{0i}$ and achieve the same marginal distribution for $Y_{i,t}$.  A common choice for a constraint is to allow the intercept of $\beta_{0i}$ to vary, but to fix one of the $\alpha$ parameters along with the $Var(Z_{i,t}) = \sigma^2_i$.  That is, to let $\bfalpha = (-\infty, \alpha_1 = 0, \alpha_2, ..., \alpha
_{J}, \infty)$ with unknown parameters $\alpha_2, ..., \alpha_{J}$ along with a fixed $Var(Z_{i,t}) = \sigma^2_i = 1$, which results in an identifiable model with no loss in generality \citep{feng2014composite, higgs2010clipped, schliep2015data, erhardt2024spatio}.  This choice --- allowing all but one element of $\bfalpha$ to vary as parameters --- achieves much flexibility in distributing the probability mass across the $J$ levels of the ordinal response, but has been shown to mix very poorly \citep{schliep2015data}.  This poor mixing adds substantially to computational cost, particularly when the number of ordinal levels $J+1$ is large.  As an alternative, here we fix $\bfalpha = (-\infty, 0, 1, ..., J-1, \infty)$, but parameterize both the intercept $\beta_{0i}$ and $Var(Z_{i,t}) = \sigma^2_i$ to allow for model flexibility. 


We define the $p^{th}$ covariate parameter for all locations as $\bfbeta_{(p)}= (\beta_{p1}, ..., \beta_{pI})', p=0,...,P$ (\textit{with} parentheses around the subscript $(p)$, to differentiate it from $\bfbeta_{i}$ which is a vector of $P+1$ covariates for location $i$ only). We define $\gamma_i \equiv \text{logit}(\rho_i)$ to be the logit-transformed temporal autoregressive parameter at location $i$. We assume each $\bfbeta_{(p)}$, $p=0,...,P$ and $\bfgamma = (\gamma_1,...,\gamma_I)$ independently follow intrinsic conditional autoregressive (ICAR) spatial models. That is, we define $a_{i,\ell}=1$ if locations $i$ and $\ell$ are neighbors, where we define a neighboring grid cell based on queen's adjacency, and $a_{i,\ell}=0$ otherwise. Let $a_{i+}=\sum_{\ell} a_{i,\ell}$ denote the total number of neighbors for location $i$. The ICAR model specified assumes
\begin{equation*}
    \gamma_i \mid \bfgamma_{-i} \sim \mathcal{N} \left(\sum_{\ell} \frac{a_{i,\ell}}{a_{i+}} \gamma_{\ell}, \frac{\sigma^2_{\gamma}}{a_{i+}} \right),
\end{equation*}
where $\bfgamma_{-i} = \{\gamma_{\ell} : \ell \neq i\}$ and $\sigma^2_{\gamma}$ is a variance parameter. This ICAR prior yields the (improper) joint density $\pi\left(\bfgamma\right) \propto \exp \left(-\frac{1}{2\sigma^2_{\gamma}} \bfgamma' (\bfD - \bfA) \bfgamma \right)$ where $\bfA$ is the symmetric adjacency matrix composed of all $a_{i,\ell}$ and $\bfD = diag(a_{1+}, ..., a_{I+})$. Note that while the ICAR is itself an improper distribution, it can be used as a prior distribution for spatial random effects \citep{banerjee2003hierarchical}. Similarly, we assume ICAR models for each regression coefficient, $\bfbeta_{(p)}$, each with variance parameter denoted by $\sigma^2_{(p)}$. All variance parameters ($\{\sigma^2_i : i=1,...,I\}, \sigma^2_{\gamma}, \{\sigma^2_{(p)} : p=0,...,P\}$) are assumed to follow weakly informative conjugate inverse gamma (IG) prior distributions. 

The autoregressive structure of $Z_{i,t}$ is assumed to capture temporal autocorrelation in drought and will assist in forecasting the drought levels in future weeks. We assume a spatial random effect for the intercept to capture the high degree of spatial autocorrelation in drought. However, we also expect the relationship between drought and the environmental variables, as well as the degree of temporal autocorrelation in drought, to vary smoothly across space. Thus, we also assume spatial random effects for the $\beta s$ and temporal autocorrelation parameters.

Define $\bfY_{1:T}$ and $\bfZ_{1:T}$ to be the $IT$-dimensional vectors of the observed ordinal and latent continuous variables, respectively, and let $\bfX_{1:T}$ denote the $IT \times (P+1)$ dimensional design matrix of covariates. Additionally, let $\bftheta_{Z,i} = \left(\bfbeta_i, \gamma_i, \sigma^2_i \right)$ for $i=1,...,I$ be the $(P+3)$-dimensional vector of site-specific parameters that appear in the model for $\bfZ_{1:T}$. Let $\bftheta_Z = \{ \bftheta_{Z,i} : i=1,...,I\}$ be the $I(P+3)$-dimensional vector the comprises the full collection of all site-specific parameters. The model also depends on hyperparameters corresponding to variances of the spatial random effect models. Let $\bfphi = \left(\sigma^2_{\gamma}, \{\sigma^2_{(p)}, p=0, ..., P\}\right)$ be the $(P+2)$-dimensional vector of spatially static hyperparameters. The full parameter vector contains all site-specific parameters in addition to the variance hyperparameters from the spatial random effect models, that is, the $(I(P+3)+P+2)$-dimensional vector $\left( \bftheta_Z, \bfphi \right)$. We can express the full Bayesian hierarchical model as the following three stages:
\begin{equation}
\begin{aligned}
        \text{Data Model: } & \bfY_{1:T} \mid \bfZ_{1:T} \sim f\left(\bfY_{1:T} \mid \bfZ_{1:T}\right)\\
        \text{Process Model: } & \bfZ_{1:T} \mid \bftheta_Z, \bfX_{1:T} \sim \pi \left( \bfZ_{1:T} \mid \bftheta_Z, \bfX_{1:T} \right) \\
        \text{Prior Models: } & \bftheta_Z, \bfphi \sim \pi\left(\bftheta_Z | \bfphi\right) \times \pi\left(\bfphi\right).
    \end{aligned}
    \label{eq:BHM}
    \end{equation}
The posterior distribution is then of the form
\begin{equation}
   \pi\left(\bfZ_{1:T}, \bftheta_Z, \bfphi \mid \bfY_{1:T}, \bfX_{1:T} \right)  \propto f\left(\bfY_{1:T} \mid \bfZ_{1:T}\right) \pi\left(\bfZ_{1:T} \mid \bftheta_Z, \bfX_{1:T}\right) \pi\left(\bftheta_Z \mid \bfphi\right) \pi\left(\bfphi\right).
   \label{eq:singlestateposterior}
\end{equation}
In theory, samples can be generated from the posterior distribution using a Gibbs sampling MCMC algorithm. However, even though the model specification permits Gibbs updates, this is computationally infeasible when the number of spatial locations and/or time periods is large. Observe that when $T$ is large, computing $\pi\left(\bfZ_{1:T} \mid \bftheta_Z, \bfX_{1:T}\right)$ is expensive, and this calculation needs to be done for each Gibbs update of the spatial random effects, which are themselves challenging to update when $I$ is big.

\subsection{Two-stage MCMC Algorithm}

To reduce the computational cost associated with fitting the model described above, we propose a two-stage MCMC approach modified from that described in \citet{lunn2013fully} to be suitable for spatio-temporal data. In the first stage, we simulate draws of the site-specific parameters from a simplified model that is computationally ``easy'' to simulate from. These draws are then used as proposals in a Metropolis-Hastings algorithm of the full model in the second stage. This alleviates computational expense associated with computing $\pi\left(\bfZ_{1:T} \mid \bftheta_Z, \bfX_{1:T}\right)$ and allows us to efficiently generate posterior samples from the full spatio-temporal model. The two stages are each described in detail below.


\subsubsection{Stage One}

The computationally efficient stage one model that we fit assumes independence across space. Specifically, we replace all ICAR priors with independent priors, $\beta_{p,i} \overset{ind}{\sim} N(0, \xi_{p}^2)$, for $p=0, ..., P$ and $i=1, ..., I$, where $\xi_{p}^2$ is known and specified such that the prior is weakly informative. In our application, we chose a standard deviation of 3 because the cutoffs of the continuous latent variable are $(0, 1, 2, 3, 4)$, which implies the intercept parameters should all fall within some range of these values, and since the covariates are all standardized we would expect the coefficients to be within some moderate range of 0. The stage one model also assumes the site-specific $\gamma_i \equiv \text{logit}(\rho_i)$ independently follow logistic distributions with mean zero and scale one, which is equivalent to $\rho_i \overset{iid}{\sim} U(0, 1)$. The prior for the variance components, $\sigma^2_i$, remains the same as the full model which assumed independent inverse gamma distributions. 

Observe that these modeling changes break all spatial dependence across locations, as the only previous dependence was through $\bfbeta_p, p=0, ..., P$ and $\bfgamma$.  These changes also temporarily remove all hyperparameters from consideration, so the full parameter vector is $\bftheta_Z$. Notably, the data model and process model given in equation \eqref{eq:BHM} are the same in the full model and stage one model, only the prior model specification differs. Let $\tilde{\pi}\left(\bftheta_Z\right)$ denote the stage one prior. The independence implies the stage one posterior distribution can be written as
\begin{equation}
        \begin{aligned}
            \tilde{\pi}\left(\bftheta_Z, \bfZ_{1:T} \mid \bfY_{1:T}, \bfX_{1:T}\right) &\propto \pi\left(\bfY_{1:T} \mid \bfZ_{1:T}\right) f\left(\bfZ_{1:T} \mid \bfX_{1:T}, \bftheta_Z\right) \tilde{\pi}\left(\bftheta_Z\right) \\
            &= \prod_{i=1}^I f\left(\bfY_{i,1:T} \mid \bfZ_{i,1:T}\right) \pi\left(\bfZ_{i,1:T} \mid \bfX_{i,1:T}, \bftheta_{Z,i}\right) \tilde{\pi}\left(\bftheta_{Z,i}\right) \\
            &\propto \prod_{i=1}^I  \tilde{\pi}\left(\bftheta_{Z,i}, \bfZ_{i,1:T} \mid \bfY_{i,1:T}, \bfX_{i,1:T}\right),
            \label{eq:stageonepost}
        \end{aligned}
\end{equation}
where $\bftheta_{Z,i} = (\bfbeta_{i}, \gamma_i, \sigma^2_i)$ is a vector of the location-specific parameters for location $i$ only.  Crucially, the stage one posterior is now a product of $I$ site-specific posterior densities, which permits the use of parallel computing to efficiently sample from the posterior by sampling each location-specific posterior in parallel. Additionally, this model specification yields full conditional distributions that are all of a known form, so the stage one posterior can be simulated with a Gibbs sampling algorithm. 
With over 3,000 locations in the application described in this paper and potentially tens or hundreds of thousands of locations in other applications, this parallelization in stage one reduces computational cost by several orders of magnitude, and the stage one computing cost becomes a relatively minor component to the overall computing cost.

\subsubsection{Stage Two MCMC}

The second stage implements a Metropolis-within-Gibbs algorithm for the \textit{full} model. The hyperparameters that only appear in the full model, $\boldsymbol{\phi}$, can be updated with Gibbs updates. The location-specific parameters that appear in both the full model and the stage one model, $\left(\bfZ_{i,1:T},  \bftheta_{Z,i}\right)$, are updated jointly for each location using a Metropolis-Hastings update, where the proposed draws are values previously generated in stage one. Let $\omega^{(m)}$ generically denote the value of a parameter $\omega$ after the $m$th iteration of the MCMC algorithm. We describe the $m$th iteration of the Stage Two MCMC algorithm below.

Without loss of generality, assume each iteration of the MCMC algorithm first updates the variance parameters in $\boldsymbol{\phi}=\left(\sigma^2_{\gamma}, \{ \sigma^2_{(p)}, p=0,...,P \} \right)$. The full conditional distribution of each is an inverse gamma distribution with shape parameter of $a + I/2$. The scale parameters are $b + 0.5\sum_{i \sim j} \left(\gamma_i^{(m-1)} - \gamma_j^{(m-1)} \right)^2$ and $b + 0.5\sum_{i \sim j} \left(\beta_{pi}^{(m-1)} - \beta_{pj}^{(m-1)} \right)^2$ for $\sigma^{2}_{\gamma}$ and $ \{ \sigma^{2}_{(p)}, p=0,...,P \}$, respectively, where $a=0.5$ and $b=0.5$ are the shape and scale parameters from the prior distribution, and $i \sim j$ is the summation over all neighbors of element $j$ as defined in the adjacency matrix $\bfA$. Let $\mathbf{\phi}^{(m)}=\left(\sigma^{2(m)}_{\gamma}, \{ \sigma^{2(m)}_{(p)}, p=0,...,P \} \right)$ denote the value of these parameters after simulating from these full conditional distributions. 

The location-specific parameters $(\bfZ_{i,1:T}, \bftheta_{Z,i})$ are then iteratively updated for $i=1,...,I$ using Metropolis-Hastings updates to simulate from their full conditional distributions. We describe the Metropolis-Hastings step here for location $i$. Observe that under the full model, all spatial dependence is specified within the model for $\bftheta_Z$, and hence the full conditional distribution for location $i$ is 
\begin{equation}
    p(\bfZ_{i,1:T}, \bftheta_{Z,i} \mid \cdot) \propto f(\bfY_{i,1:T} \mid \bfZ_{i,1:T}) \pi(\bfZ_{i,1:T} \mid \bftheta_{Z,i}, \bfX_{i,1:T}) \pi(\bftheta_{Z,i} \mid \bftheta_{Z,-i}, \bfphi ),
    \label{eq:fullpost}
\end{equation}
where $$\pi(\bftheta_{Z,i} \mid \bftheta_{Z,-i}, \bfphi ) = \pi(\gamma_i \mid \bfgamma_{-i}, \sigma^2_{\gamma}) \prod_{p=0}^P \pi(\beta_{pi} \mid \bfbeta_{p,-i}, \sigma^2_{(p)})$$ is the product of the ICAR conditional distributions for each of the spatial random effects. 

 Let $\left(\bfZ_{j,1:T}^{(m^*)}, \bftheta_j^{(m^*)} \right)  = \left(\bfZ_{j,1:T}^{(m^*)}, \bfbeta_j^{(m^*)}, \gamma_j^{(m^*)}, \sigma^{2(m^*)}_j \right)$ denote the current value of the site-specific parameters for location $j$ during iteration $m$ of the MCMC algorithm. Since we update each location iteratively, that means $m^* = m$ for $j=1,...,i-1$ and $m^* = m-1$ for locations $j=i,...,I$ which have not yet been updated.  To propose a new parameter vector for location $i$, we randomly sample from the stage one posterior. More specifically, we randomly sample $\left(\bfZ_{i,1:T}^*, \bftheta_i^*\right) = \left(\bfZ_{i,1:T}^*,\bfbeta_i^*, \gamma_i^*, \sigma^{2*}_i \right)$ from the stage one MCMC output. The independence of the stage one posterior shown in equation \eqref{eq:stageonepost} implies the proposal distribution is
 \begin{equation}
 \begin{aligned}
     q\left(\bfZ_{i,1:T}^*, \bftheta_i^* \right) &\propto  \tilde{\pi}\left(\bftheta_{Z,i}^*, \bfZ_{i,1:T}^* \mid \bfY_{i,1:T}, \bfX_{i,1:T}\right) \\
     &= f(\bfY_{i,1:T} \mid \bfZ_{i,1:T}^*) \pi(\bfZ_{i,1:T}^* \mid \bftheta_{Z,i}^*, \bfX_{i,1:T}) \tilde{\pi}(\bftheta_{Z,i}^*)
     \label{eq:proposal}
\end{aligned}
 \end{equation}
The proposed sample is accepted and $\left(\bfZ_{i,1:T}^{(m)}, \bftheta_i^{(m)} \right) = \left(\bfZ_{i,1:T}^*, \bftheta_i^*\right)$ with probability $R' = \min(1,R)$, where the acceptance ratio $R$ depends on equations \eqref{eq:fullpost} and \eqref{eq:proposal} such that
\begin{equation}
    \begin{aligned}
        R &= \frac{p\left(\bfZ_{i,1:T}^*, \bftheta_{Z,i}^* \mid \cdot\right)}{p\left(\bfZ_{i,1:T}^{(m-1)}, \bftheta_{Z,i}^{(m-1)} \mid \cdot \right)} \times \frac{q\left (\bfZ_{i,1:T}^{(m-1)}, \bftheta_i^{(m-1)}\right)}{q\left (\bfZ_{i,1:T}^{*}, \bftheta_i^{*}\right)} \\
        &=\frac{f(\bfY_{i,1:T} \mid \bfZ_{i,1:T}^*) \pi(\bfZ_{i,1:T}^* \mid \bftheta_{Z,i}^*, \bfX_{i,1:T}) \pi(\bftheta_{Z,i}^* \mid \bftheta_{Z,-i}^{(m^*)}, \bfphi^{(m)} )}{f(\bfY_{i,1:T} \mid \bfZ_{i,1:T}^{(m-1)}) \pi(\bfZ_{i,1:T}^{(m-1)} \mid \bftheta_{Z,i}^{(m-1)}, \bfX_{i,1:T}) \pi(\bftheta_{Z,i}^{(m-1)} \mid \bftheta_{Z,-i}^{(m^*)}, \bfphi^{(m)} )} \\
        &\times \frac{ f(\bfY_{i,1:T} \mid \bfZ_{i,1:T}^{(m-1)}) \pi(\bfZ_{i,1:T}^{(m-1)} \mid \bftheta_{Z,i}^{(m-1)}, \bfX_{i,1:T}) \tilde{\pi}(\bftheta_{Z,i}^{(m-1)})}{ f(\bfY_{i,1:T} \mid \bfZ_{i,1:T}^*) \pi(\bfZ_{i,1:T}^* \mid \bftheta_{Z,i}^*, \bfX_{i,1:T}) \tilde{\pi}(\bftheta_{Z,i}^*)} \\
        &= \frac{\pi(\bftheta_{Z,i}^* \mid \bftheta_{Z,-i}^{(m^*)}, \bfphi^{(m)} ) \tilde{\pi}(\bftheta_{Z,i}^{(m-1)})}{ \pi(\bftheta_{Z,i}^{(m-1)} \mid \bftheta_{Z,-i}^{(m^*)}, \bfphi^{(m)} ) \tilde{\pi}(\bftheta_{Z,i}^*)}.
    \end{aligned}
    \label{eq:R0}
\end{equation}

The value of using the stage one posterior as the proposal density for stage two is apparent with the drastic simplification it affords $R$. Notably, the acceptance ratio $R$ does not depend on $ \pi(\bfZ_{i,1:T} \mid \bftheta_{Z,i}, \bfX_{i,1:T})$, which alleviates the computational burden since evaluation of this density is expensive when $T$ is large as it is in our application, and it instead only depends on the process-level models for $\bftheta_{Z,i}$. Under our application model specification, equation \eqref{eq:R0} becomes
\begin{equation}
    \begin{aligned}
         R &= \frac{ \pi(\gamma_i^* \mid \bfgamma_{-i}^{(m^*)}, \sigma^{2(m)}_{\gamma}) \prod_{p=0}^P \pi(\beta_{pi}^* \mid \bfbeta_{p,-i}^{(m^*)}, \sigma^{2(m)}_{(p)})}{ \pi(\gamma_i^{(m-1)} \mid \bfgamma_{-i}^{(m^*)}, \sigma^{2(m)}_{\gamma}) \prod_{p=0}^P \pi(\beta_{pi}^{(m-1)} \mid \bfbeta_{p,-i}^{(m^*)}, \sigma^{2(m)}_{(p)})} \frac{ \tilde{\pi}(\gamma_i^{(m-1)}) \prod_{p=0}^P \tilde{\pi}(\beta_{pi}^{(m-1)})}{\tilde{\pi}(\gamma_i^{*}) \prod_{p=0}^P \tilde{\pi}(\beta_{pi}^{*})} \\
         &= \frac{\exp\left(-\frac{w_{i+}}{2\sigma^{2(m)}}\left(\gamma_i^* - \bar{\bfgamma}_{j \sim i}^{(m^*)}\right)^2 \right) }{\exp\left(-\frac{w_{i+}}{2\sigma^{2(m)}}\left(\gamma_i^{(m-1)} - \bar{\bfgamma}_{j \sim i}^{(m^*)}\right)^2 \right)}  \frac{\exp\left( -\gamma_i^{(m-1)} \right) \left(1+\exp\left( -\gamma_i^{*} \right) \right)^2}{\left(1+\exp\left( -\gamma_i^{(m-1)} \right) \right)^2 \exp\left( -\gamma_i^* \right) }   \\
         &\times  \frac{\prod_{p=0}^P \exp\left(-\frac{w_{i+}}{2\sigma_{(p)}^{2(m)}}\left(\beta_{pi}^* - \bar{\bfbeta}_{p,j \sim i}^{(m^*)}\right)^2 \right) }{\prod_{p=0}^P  \exp\left(-\frac{w_{i+}}{2\sigma_{(p)}^{2(m)}}\left(\beta_{pi}^{(m-1)} - \bar{\bfbeta}_{p,j \sim i}^{(m^*)}\right)^2 \right) } \frac{\prod_{p=0}^P \exp\left(-\frac{1}{2\xi_p^2}\left(\beta_{pi}^{(m-1)}\right)^2\right) }{\prod_{p=0}^P \exp\left(-\frac{1}{2\xi_p^2}\left(\beta_{pi}^{*}\right)^2\right)},
    \end{aligned}
    \label{eq:R}
\end{equation}
where $j \sim i$ refers to the neighbors of grid cell $i$, and $\bar{\gamma}_{j \sim i}$ and $\bar{\beta}_{j \sim i}$ refers to the means of each quantity taken over the neighboring grid cells only.  So, the acceptance ratio only depends on the conditional distributions specified for the $P+2$ site-specific regression and autocorrelation parameters in the full model (ICAR models) and the stage one model (independent normal and logistic models). This process is repeated iteratively for $i=1,...,I$ within each iteration of the MCMC algorithm.

\section{Comparison of Single Stage and Two Stage Methodologies}
\label{se:comparison}

Here we compare the computational efficiency of our two-stage methodology to a standard single-stage Metropolis-within-Gibbs MCMC approach.  To clarify terminology, we call the \textit{single-stage MCMC} the algorithm which samples directly from equation \eqref{eq:singlestateposterior}, which is computationally infeasible for large datasets.  We call \textit{stage one} of the \textit{two-stage} algorithm sampling from equation \eqref{eq:stageonepost}, which is done independently and in parallel.  We call \textit{stage two} of the \textit{two-stage} algorithm the re-sampling using the acceptance ratio $R$ from equation \eqref{eq:R}. As the single-stage MCMC algorithm cannot be implemented on our full data set due to the computational cost, the comparison is done for a subset of our data.  Specifically, here we consider only locations west of 105 degree longitude and weeks between January 2020 through March 2022.  This yields $I = 1198$ locations and $T = 117$ weeks.

We implemented both a single-stage MCMC as well as our two-stage method.  Specifically, for the single-stage MCMC we ran a standard Metropolis-within-Gibbs algorithm using NIMBLE \citep{de2017programming}, which allows users to build hierarchical models using the BUGS language and objects in R, but compiles and runs these models in the often much faster C++ language.  We ran the MCMC for 45,000 iterations, using a burn-in of 20,000 and thinning by storing every 5th iteration after the burn-in.  We used automated factor slice sampling for the vector of each $\bfZ_{i,1:T}$ \citep{tibbits2014automated, turek2017automated}.  We also implemented our two-stage MCMC as described in Section 3, running the site-specific stage one runs in NIMBLE and then coding the second stage in R. Each location-specific stage one MCMC was performed for 100,000 iterations with a burn-in of 20,000, with the output thinned by only storing every 8th iteration. Notably, these were run in parallel for each location by submitting an array job.  Computations were performed using the Wake Forest University (WFU) High Performance Computing Facility\citep{WakeHPC}.  Accordingly, all $I=1198$ Markov chains were run simultaneously. The stage two MCMC was performed for 45,000 with a burn-in of 20,000 iterations and thinned by every 5th iteration, so that the total number of iterations was the same as in the single-stage approach, and both approaches have total final sample size of 5000.

\begin{figure}
    \centering
    \subfigure[]{\includegraphics[width=.32\textwidth, height=3.7in]{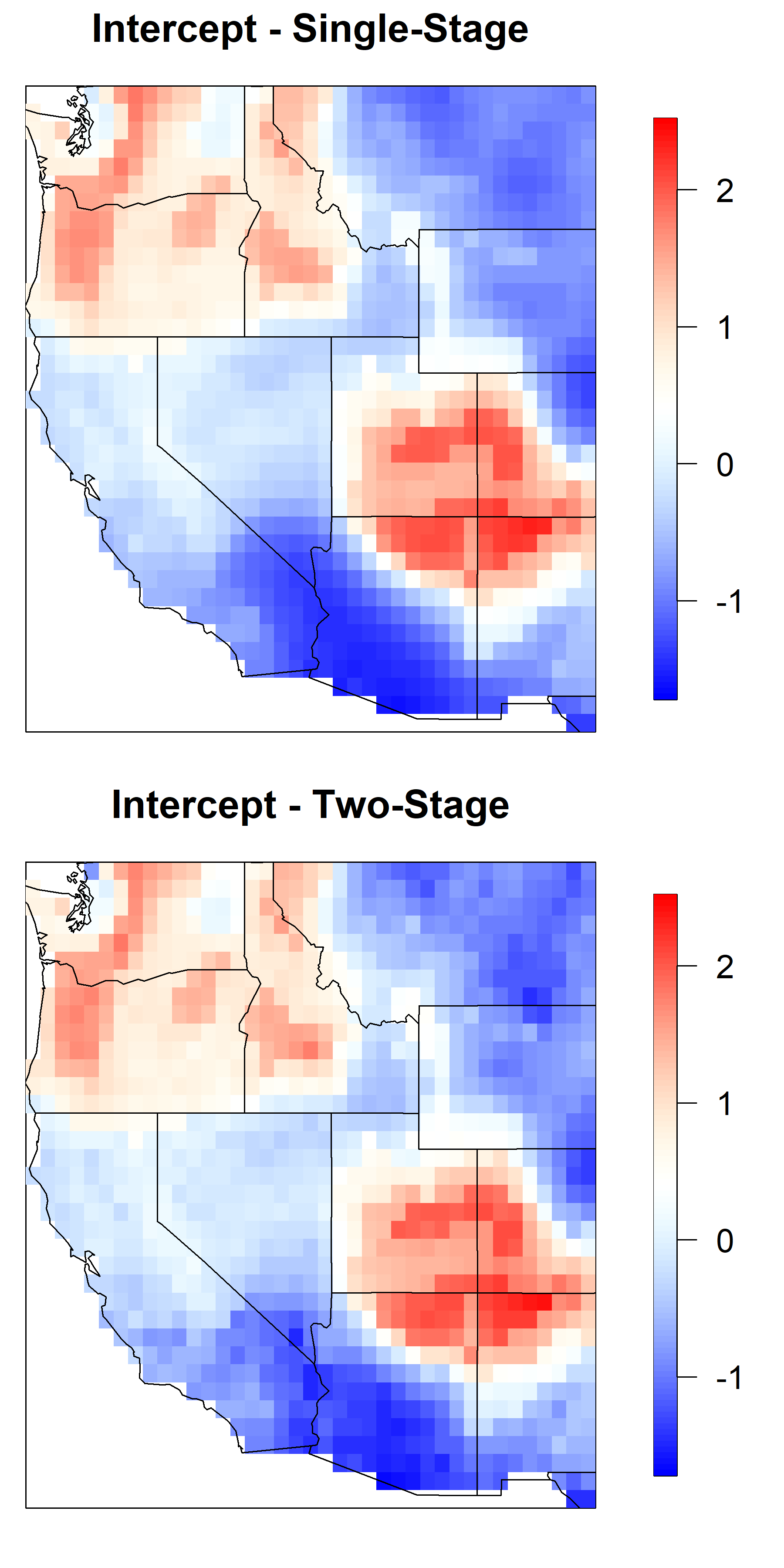}}
    \subfigure[]{\includegraphics[width=.32\textwidth, height=3.7in]{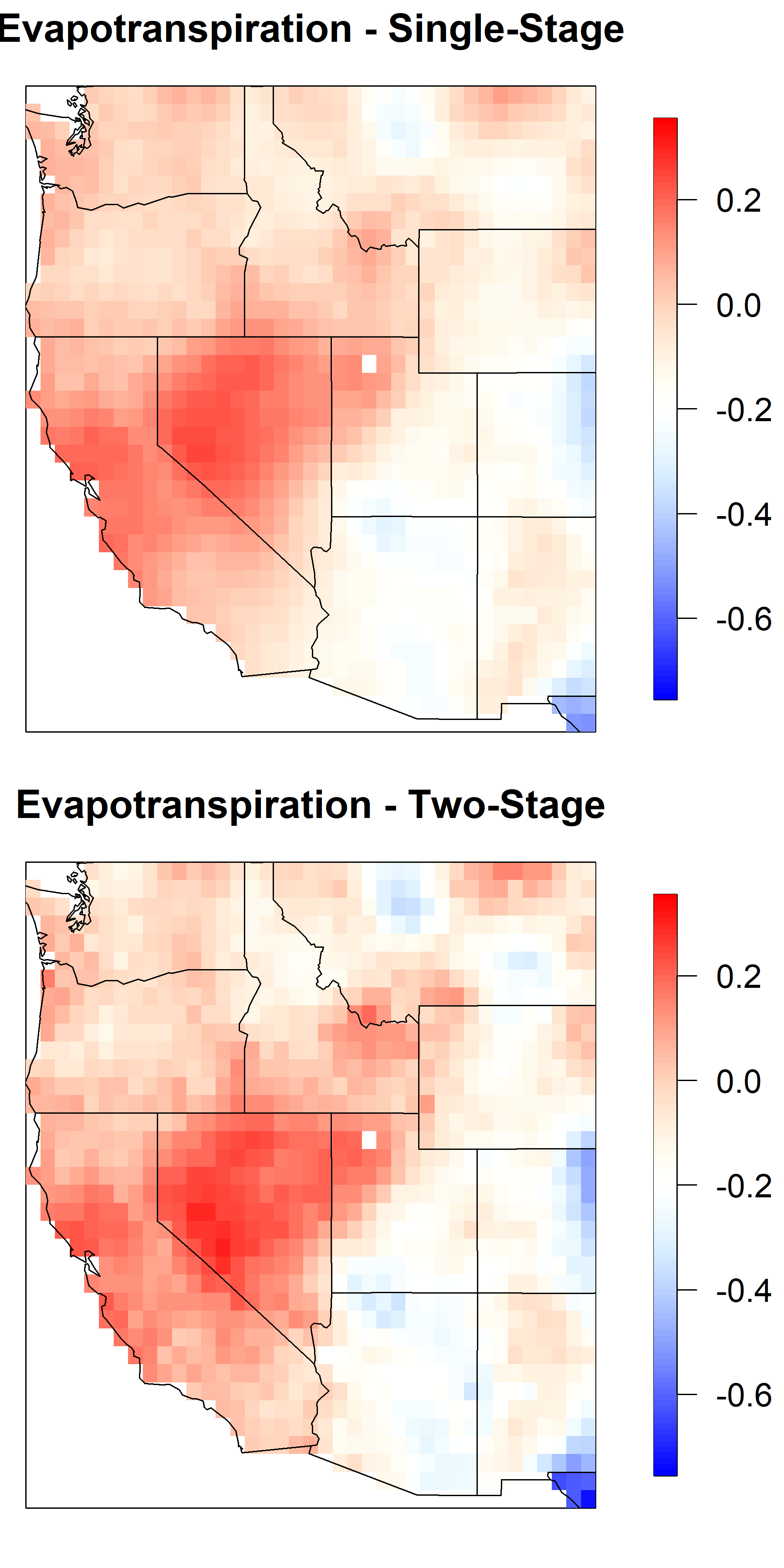}}
    \subfigure[]{\includegraphics[width=.32\textwidth, height=3.7in]{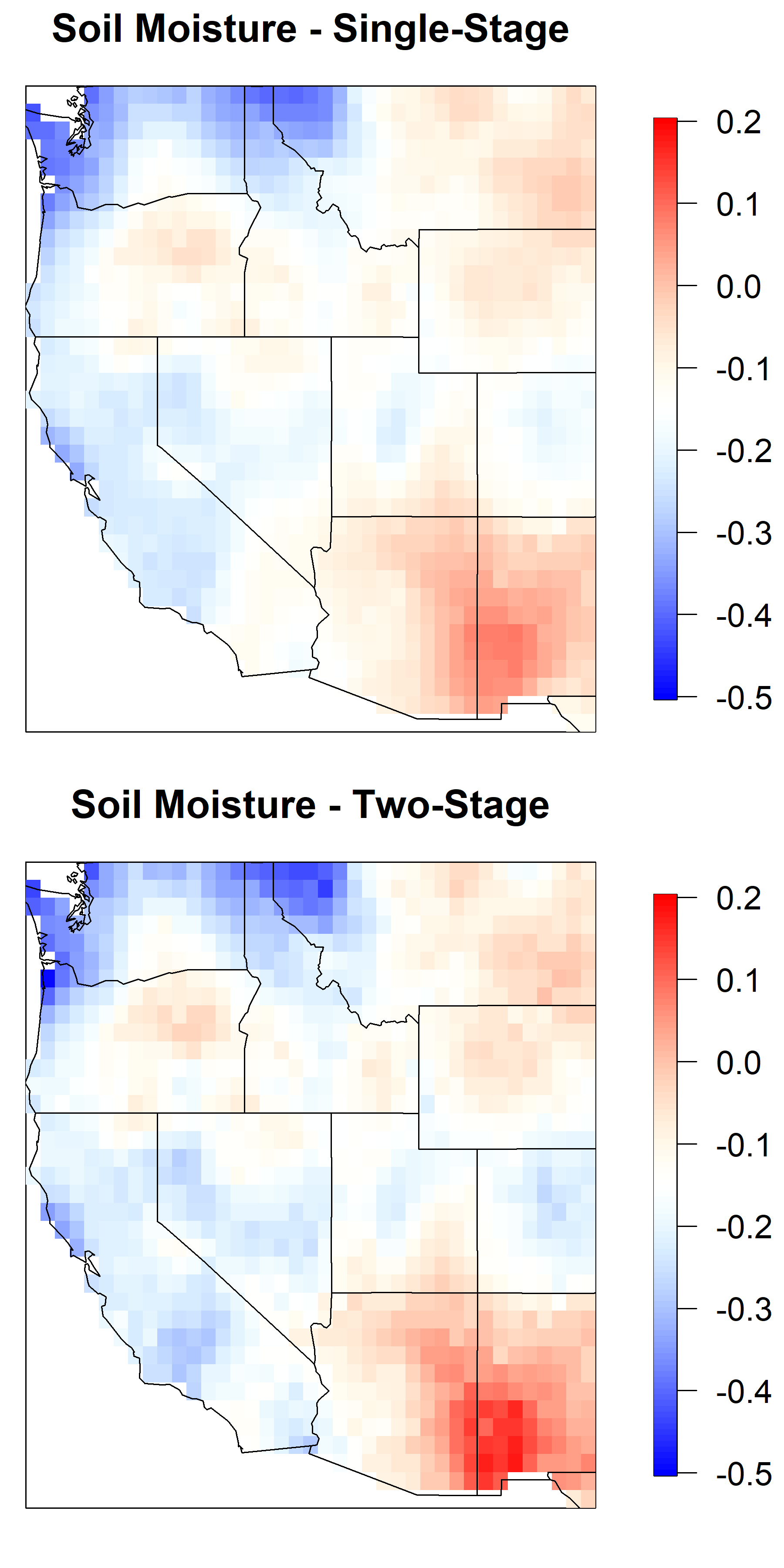}}
    \subfigure[]{\includegraphics[width=.32\textwidth, height=3.7in]{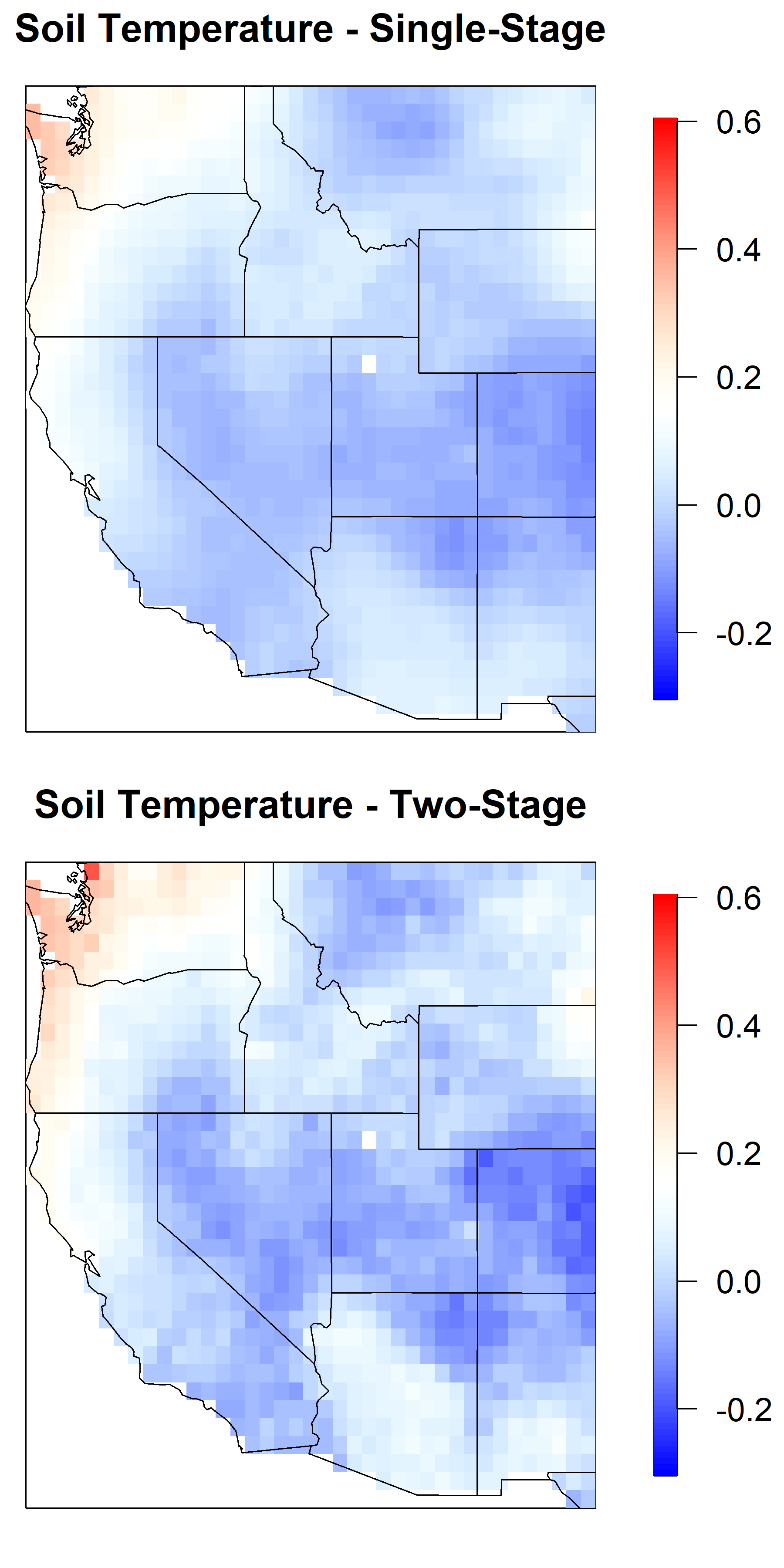}}
    \subfigure[]{\includegraphics[width=.32\textwidth, height=3.7in]{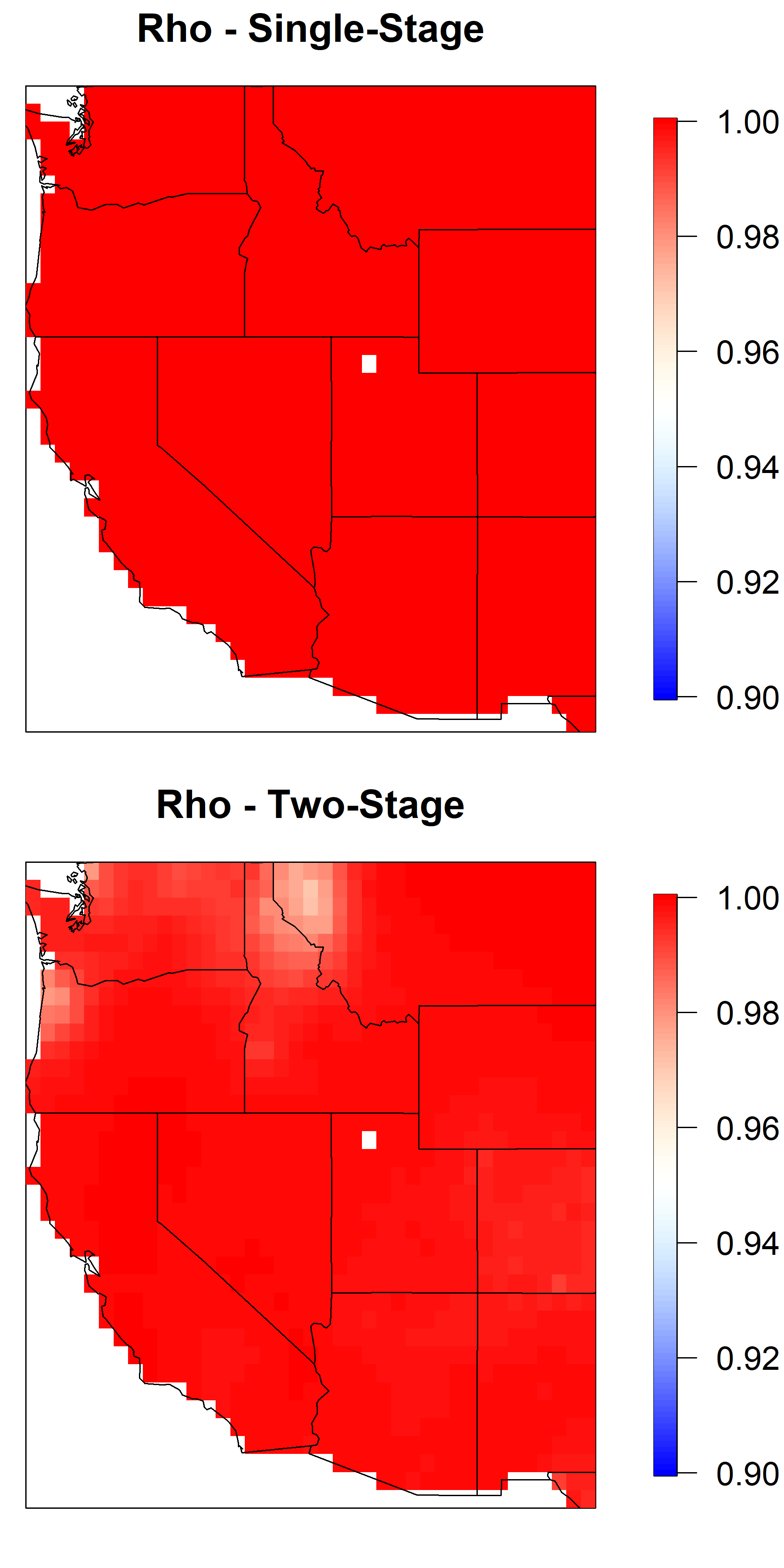}}
    \subfigure[]{\includegraphics[width=.32\textwidth, height=3.7in]{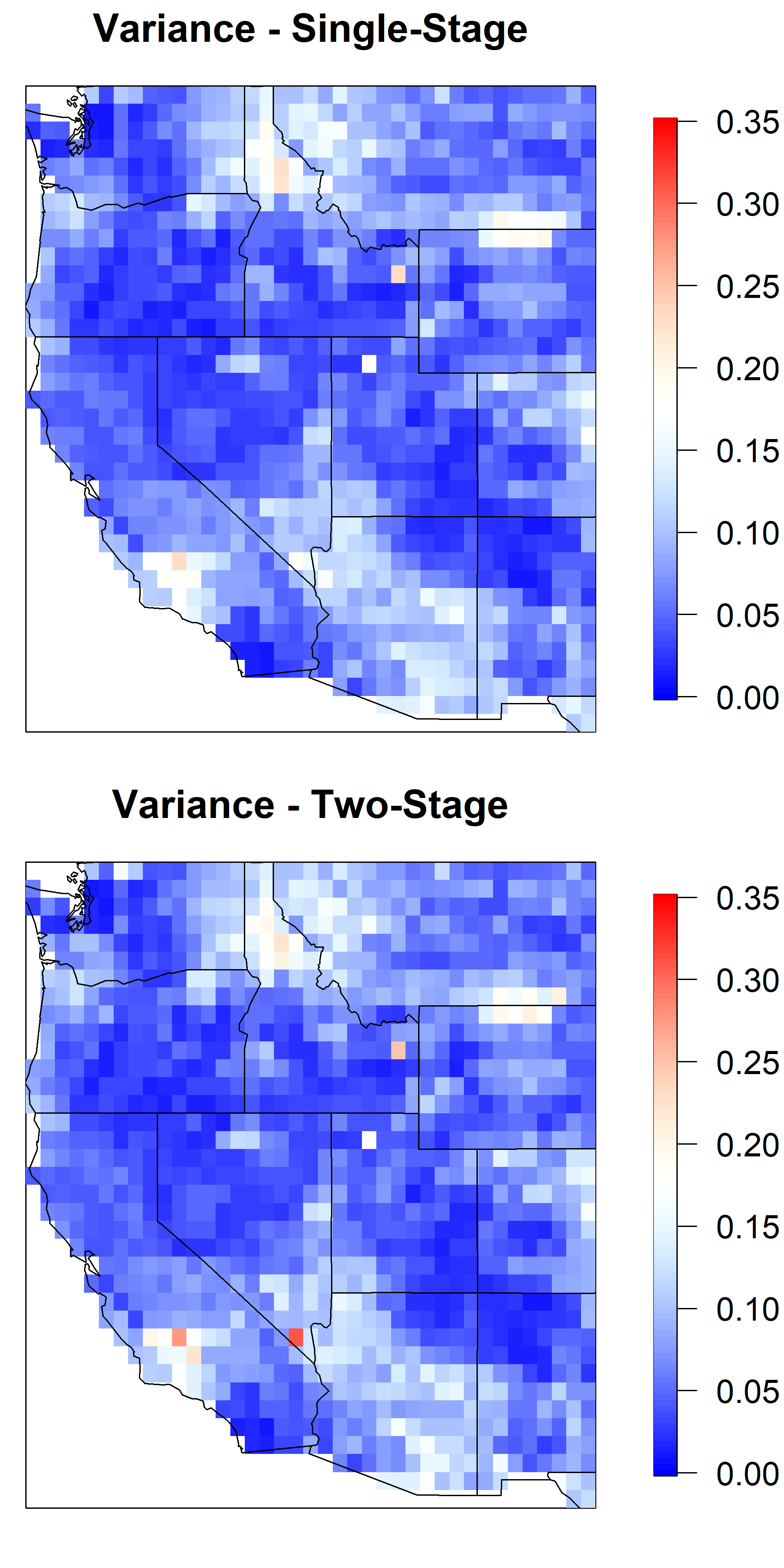}}
    \caption{Posterior means of the regression coefficients (panels (a)-(d)), temporal autocorrelation parameter (panel (e)), and variance parameter (panel (f)) for the single-stage MCMC (top row) and two-stage MCMC (bottom row). }
    \label{fig:singlevstwo}
\end{figure}

Figure \ref{fig:singlevstwo} shows maps of the posterior means of all parameters in $\bftheta_Z$ after implementing both the single-stage and two-stage MCMC algorithms. Each parameter's posterior mean is shown twice, once for the single-stage (top) and once for the two-stage (bottom).  Observe that the posterior means are nearly identical for all parameters shown --- the intercept, three regression coefficients, autoregression parameter, and variance parameter.  This is expected, as both MCMC algorithms have the same stationary distribution.  The corresponding figure of standard deviations is included in the appendix as figure \ref{fig:singlevstwoSD}.

Given the similarity in posteriors produced in the single-stage and two-stage algorithms, we turn next to the comparisons of computing time to highlight the advantages of the two-stage algorithm. Table \ref{tab:computetime1} shows the total compute time as well as the computing time required per 1000 effective samples, where the effective sample size (ESS) was computed using the \texttt{coda} R package and was averaged over all regression coefficients. For the single stage model, the computing time includes the compile and the MCMC run time. For this implementation, the compiling time was 18.42 hours and the MCMC run time was 97.38 hours, for a total compute time of 115.8 hours. We define the compute time per 1000 effective samples as the time \textit{post-compiling}, which for the single stage method yields 47.69 hours. For our proposed two-stage approach, the total time includes the time required to perform both stage one and stage two. For this illustration, the stage one MCMC took 0.17 hours and stage two took 0.78 hours, which totaled 0.95 hours. The time to obtain 1000 effective samples was 10.6 hours. From table \ref{tab:computetime1}, we see that our proposed two-stage MCMC algorithm has a drastic impact on the overall compute time, with the total time just 0.82\% of the compute time for the standard single-stage approach. While the proposed two-stage algorithm has a drastic impact on overall compute time, it is less efficient due to replacing what would be Gibbs updates in the single-stage approach with a Metropolis-Hastings update in the second stage of the two-stage approach. Even so, the proposed two-stage algorithm still has a drastic reduction in the time required to obtain 1000 effective samples, with the required 10.6 hours being less than a quarter of the time required for the standard single-stage MCMC. 

\begin{table}[!h]
    \centering
    \begin{tabular}{c|c|c}
Model & Total Time & Time to 1000 effective samples\\
\hline
        Single stage spatial model & 115.80 &  47.69 \\
        Two-stage spatial model & 0.95  &  10.6 \\
    \end{tabular}
    \caption{Computing times (in hours) for the single-stage and two-stage algorithms.  Times shown include: total time to run MCMC, which for the single-stage model includes compile and run time, and for the two-stage model includes times for both stages; and time to complete 1000 effective samples from the MCMC (averaged over all of the $\beta$ parameters), which does not include compile time for the single stage model.}
    \label{tab:computetime1}
\end{table}

Table \ref{tab:ESS1} shows the ESS for the single-stage and two-stage approaches for each of the location-specific parameters in $\bftheta_{Z,i}$. The ESS has been averaged over all locations for each model. This table also shows the ESS per hour of run time, which is the ESS dividing by the computing time required to run the MCMC. Note that for the single-stage approach, this only includes the run time and not the NIMBLE compile time. This table further depicts the fact that the standard Metropolis-within-Gibbs algorithm is more efficient, as illustrated by the much higher ESS values for the regression coefficients which are updated with Gibbs sampling updates in the single-stage approach. However, the ESS per hour is substantially greater for the two-stage algorithm. The benefit of the two-stage algorithm would be even more dramatic if we had included compile time. 

Taken together, this comparison study demonstrates that the two-stage algorithm is able to produce a substantially similar posterior distribution at a fraction of the computational cost.  In the next section, we show an application of the two-stage algorithm on our full data.  For this application, the single-stage algorithm is not computationally possible.

\begin{table}[!h]
    \centering
    \begin{tabular}{c|c|c|c|c}
    & \multicolumn{2}{c|}{Single-stage model} & \multicolumn{2}{c}{Two-stage model}\\
    \hline
    Parameter & ESS & ESS per hour & ESS & ESS per hour\\
    \hline
        $\beta_0$ & 1964.0 & 20.2 & 102.9 & 107.5 \\
        $\beta_{evp}$ & 2114.3 & 21.7 & 90.3 & 94.5 \\
        $\beta_{soilm}$ & 2163.2 & 22.2 & 88.9 & 92.9 \\
        $\beta_{tsoil}$ & 1943.9 & 20.0 & 81.4 & 85.1 \\
        $\rho$ & 208.2 & 2.1  & 92.1 & 96.2 \\
        $\sigma^2$ & 205.4 & 2.1 & 100.0 & 104.5 \\
    \end{tabular}
    \caption{Effective sample sizes and effective sample sizes per hour of computing time for location-specific parameters. Note these are averaged over all locations, and are run times only which do not include compile time.}
    \label{tab:ESS1}
\end{table}

%

\section{Application}
\label{se:application}


\subsection{Model Results}
Here we demonstrate our two-stage MCMC algorithm using the full data set, with all $I = 3254$ locations and $T=587$ time periods (we leave out the final 13 weeks as a test set for forecasting described later). The stage one MCMC algorithms were implemented in parallel as an array job on the DEAC high performance computing cluster \citep{WakeHPC}. Each site-specific stage one MCMC was run using NIMBLE for 100,000 total iterations, using a burn-in of 20,000 iterations and thinning the output by storing every 16th iteration. The stage two process was implemented on a single core of the DEAC cluster for 55,000 total iterations, with a burn-in of 5,000 iterations and thinning the output by storing every 10th iteration.

\begin{figure}
    \centering
    \subfigure[Posterior mean of intercept $\beta_0$]
    {\includegraphics[trim={1cm 1cm 0 0cm},clip, page=1,width=0.49\textwidth, height=5.5cm]{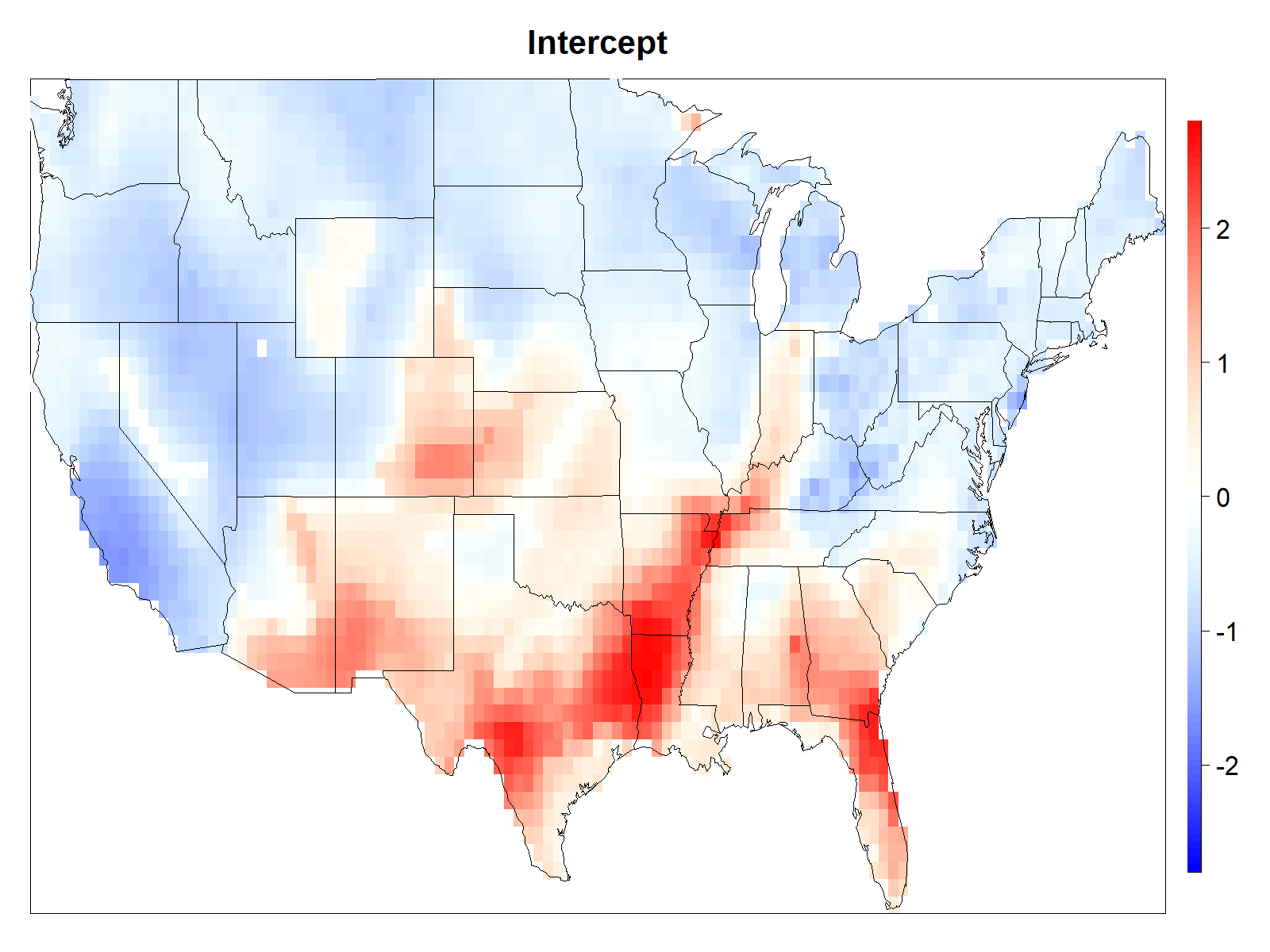}}  
    \subfigure[Posterior mean of evapotranspiration $\beta_1$]{\includegraphics[trim={1cm 1cm 0 0cm},clip,page=3,width=0.49\textwidth, height=5.5cm]{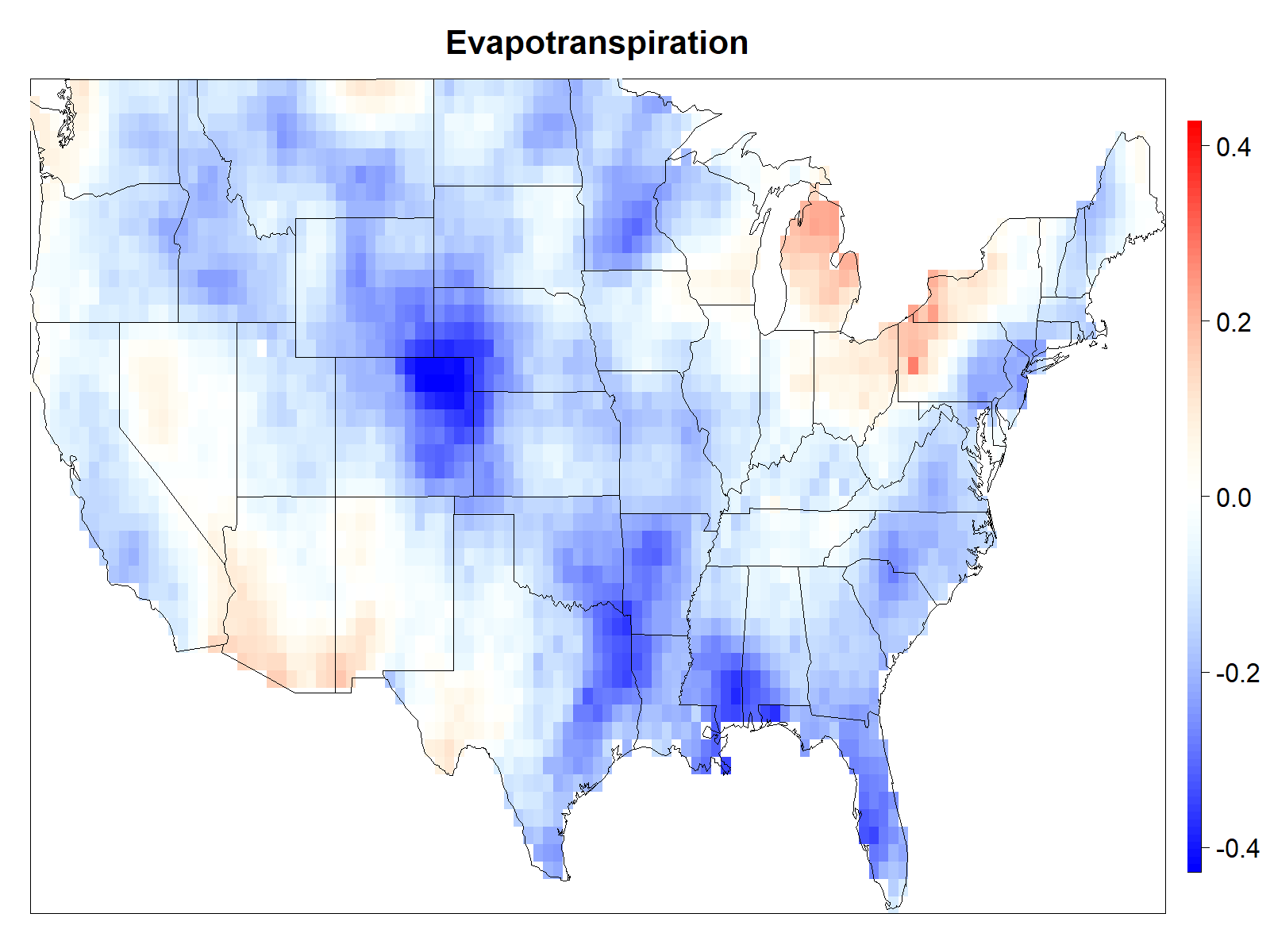}} 
    \subfigure[Posterior mean of soil moisture $\beta_2$]{\includegraphics[trim={1cm 1cm 0 0cm},clip,page=5,width=0.49\textwidth, height=5.5cm]{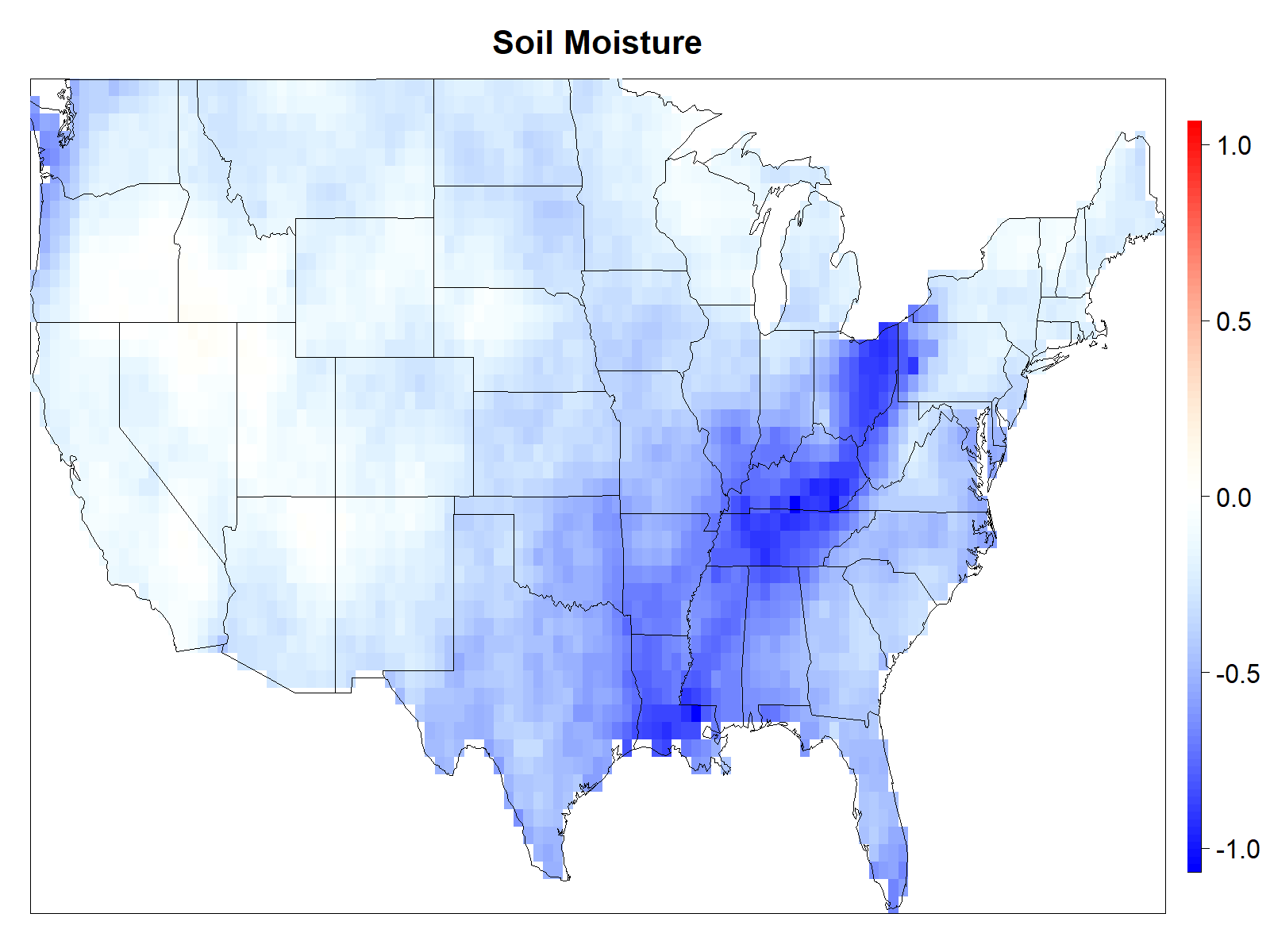}}
    \subfigure[Posterior mean of soil temperature $\beta_3$]{\includegraphics[trim={1cm 1cm 0 0cm},clip,page=7,width=0.49\textwidth, height=5.5cm]{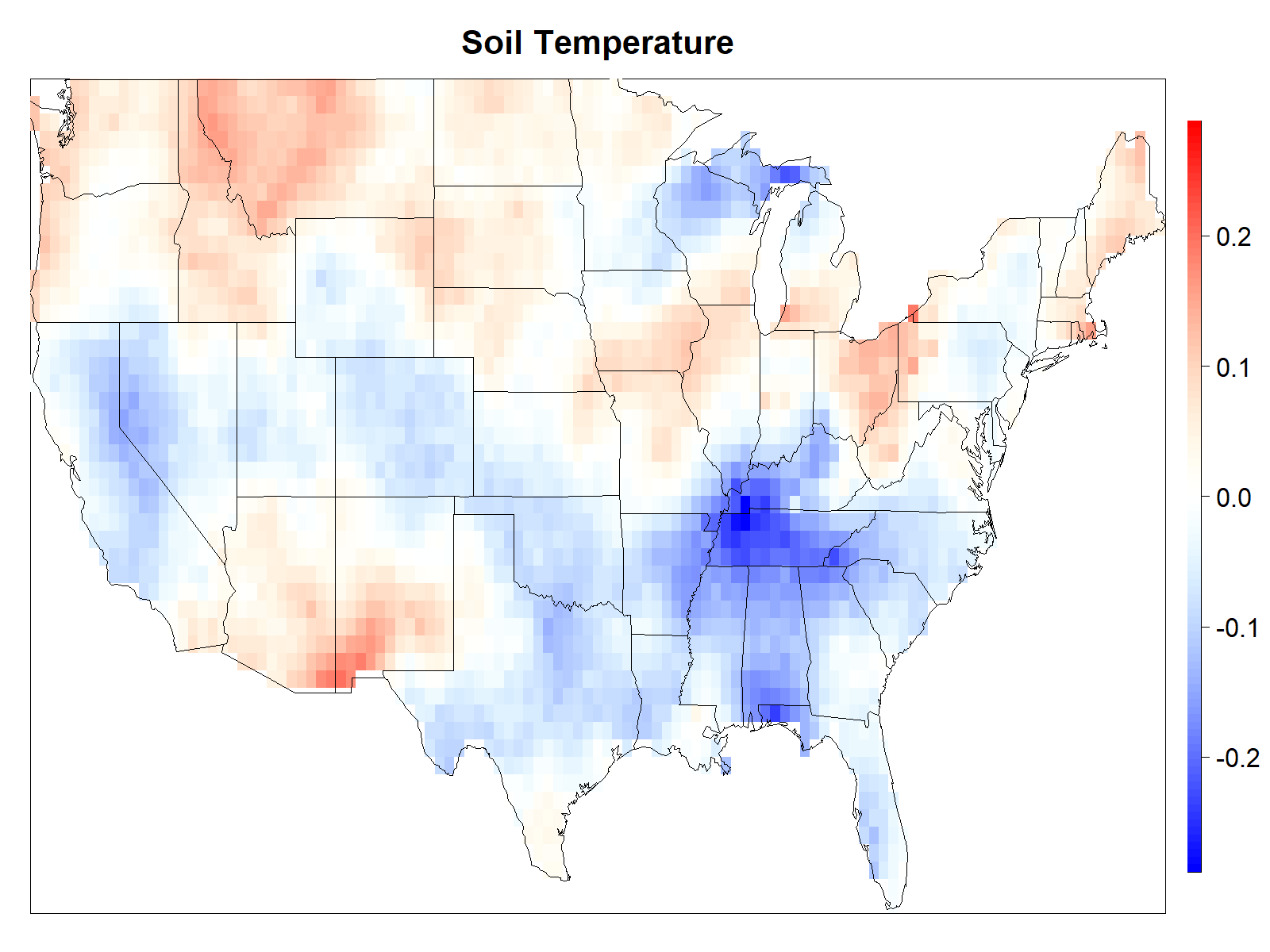}}
    \subfigure[Posterior mean of variance $\sigma^2$]{\includegraphics[trim={1cm 1cm 0 0cm},clip,page=11,width=0.49\textwidth, height=5.5cm]{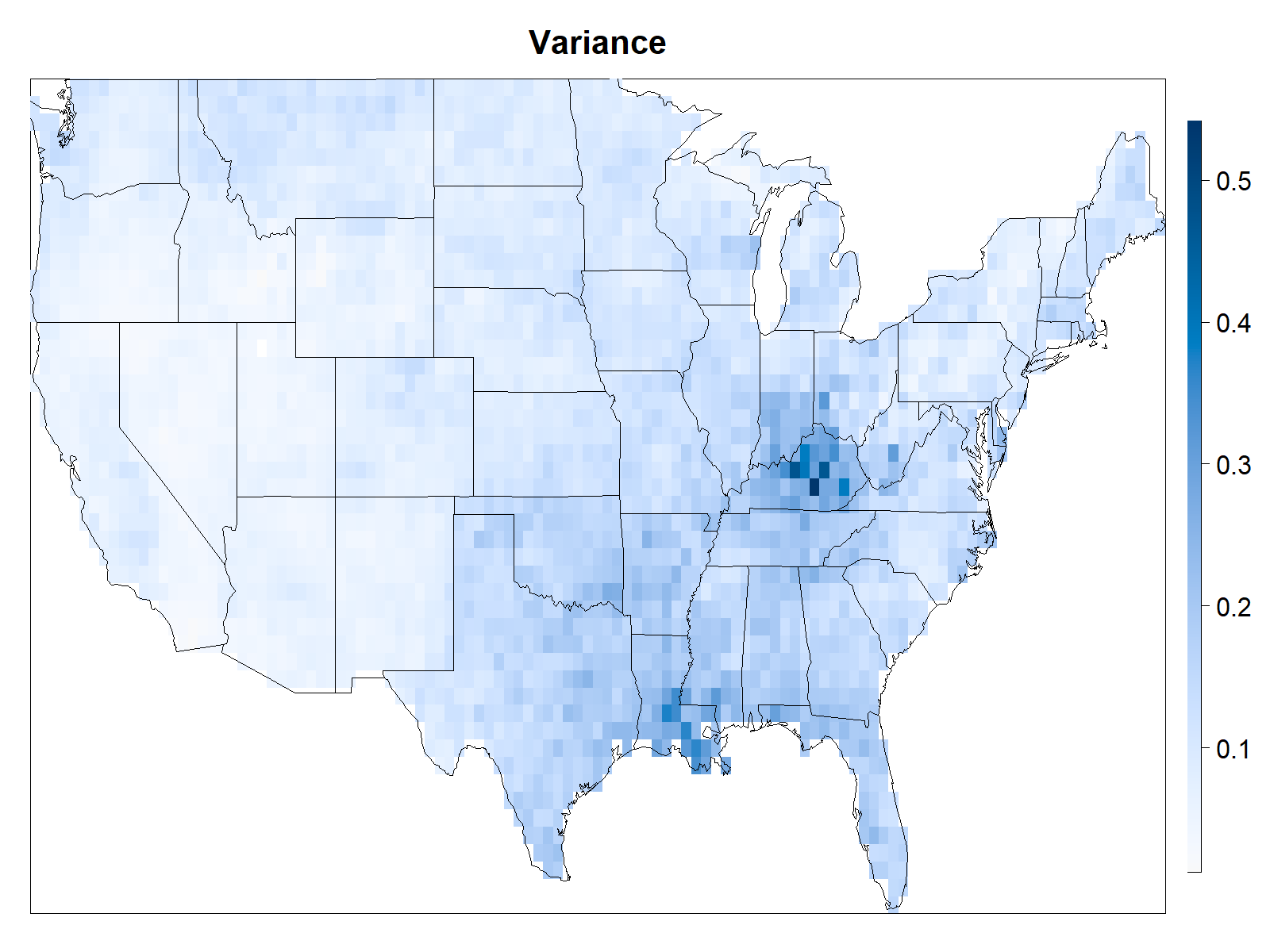}} 
    \subfigure[Posterior mean of autoregression $\rho$]{\includegraphics[trim={1cm 1cm 0 0cm},clip,page=9,width=0.49\textwidth, height=5.5cm]{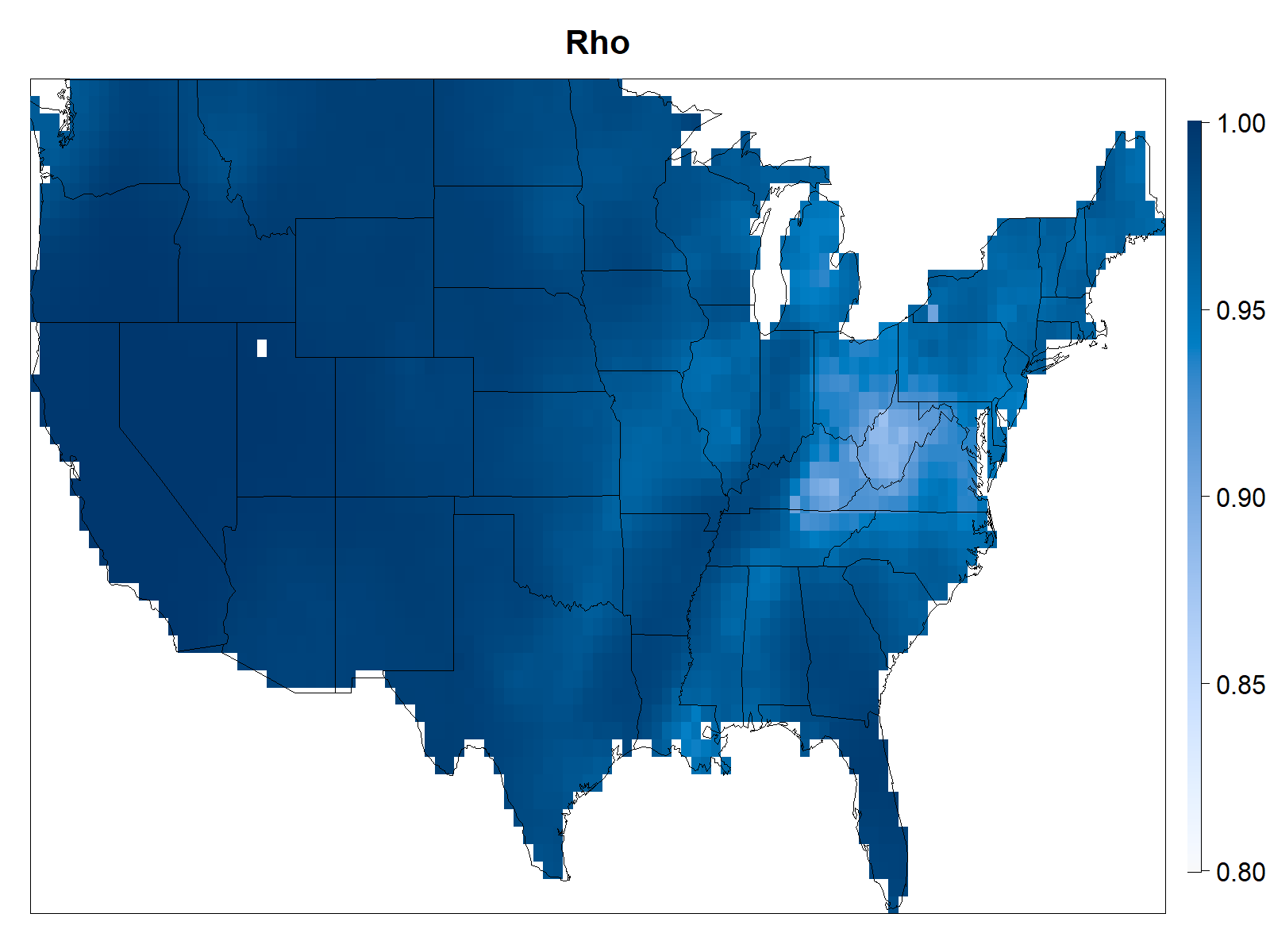}} 

    \caption{Posterior means for the regression coefficients taken from the two-stage model, implemented for the US.  Parameters shown include: (a) intercept $\beta_0$; (b) effect of evapotranspiration $\beta_1$; (c) effect of soil moisture $\beta_{2}$; (d) effect of soil temperature $\beta_{3}$; (e) variance $\sigma^2$;  and (f) autoregression $\rho$.}.
    \label{fig:betaest}
\end{figure}

Figure \ref{fig:betaest} shows posterior means for the intercept $\beta_0$, three covariate effects for evapotranspiration, soil moisture, and soil temperature, and the variance and temporal autoregression parameters. Posterior standard deviations for these are shown in the Appendix in figure \ref{fig:betasd}.  For the covariate effects, panel (c) in figure \ref{fig:betaest} shows the overall negative relationship between soil moisture and drought, with the strongest effect observed to the west of the Appalachian mountain range through the midwestern and southern United States.  The spatial variability in the variance $\sigma^2_i$ allows for model flexibility to assign probability mass across the six levels of drought differently for different regions of the United States.    The autocorrelation parameter $\rho_i$ in panel (f) demonstrates the very high predictability of the latent process $Z_{i,t}$.  This predictability enables forecasting of drought by sampling from the posterior predictive distribution.  This is demonstrated below in Section 5.2. 




\subsection{Forecasting Drought}

In our application on drought, a central scientific question is to forecast drought levels with uncertainty.  This is easily achieved by sampling from the posterior predictive distribution $p(\bfY_{T+1:T+\ell} \mid \bfY_{1:T})$, where $T$ is the end of the observed data record and $\ell$ is the number of time periods into the future where one wishes to forecast. These forecasts require covariates $\bfX_{T+1:T+\ell}$ for the future time periods, which are unavailable. To obtain these, we add a layer to the process model in equation \eqref{eq:BHM} for the covariates. The covariate model can be fit independently of the model for observed drought using a similar two-stage process, and then output from the two models is concatenated to generate samples from the posterior predictive distribution. The covariate model is described below.

As before, $\bfX_{i,t}$ is the $J$-dimensional vector of covariates measured at location $i$ and time period $t$. Prior to modeling, we detrend the covariates by subtracting the site-specific trend obtained by fitting a linear model using sets of Fourier coefficients to account for seasonality. More specifically, let $\tilde{X}_{i,t,j} = X_{i,t,j}-\hat{\mu}_{i,t,j}$ where $\hat{\mu}_{i,t,j}$ is the estimated least squares line fitting a linear regression model with mean
\begin{equation} 
\mu_{i,t,j} = \zeta_{0,i,j} + \sum_{k=1}^5 \zeta_{k,i,j}^{(s)} \text{sin}\left(\frac{2\pi k t}{365/7}  \right)+ \sum_{k=1}^5 \zeta_{k,i,j}^{(c)} \text{cos}\left(\frac{2\pi k t}{365/7}  \right).
\label{eq:ts}
\end{equation}

We then assume the vector of detrended covariates $\tilde{\bfX}_{i,t}$ follows a spatial vector autoregressive model given by 
        \begin{equation} 
\tilde{\bfX}_{i,t} = \begin{cases} 
 \bfomega_{i,t} & \text{ for } t=1 
\\
\Delta_i \tilde{\bfX}_{i, t-1}  + \bfomega_{i,t}
 & \text{ for } t > 1 ,
 \end{cases}
 \label{eq:covAR}
\end{equation}
where $\bfomega_{i,t}\overset{ind}{\sim}MVN_J(\mathbf{0},\Sigma_i)$ and $\Delta_i = \text{diag}(\delta_{1,i},...,\delta_{J,i})$. 
The autoregressive parameters $\delta_{j,i}, j=1, ..., J; i=1, ..., I$ are spatially varying and are assumed to have intrinsic conditional autoregressive prior distributions that are a priori independent for each parameter $j$. The $J \times J$ covariance matrices $\Sigma_i$ capture dependence across the $J$ covariates, and have inverse Wishart prior distributions with $J$ degrees of freedom and identity scale matrix. The remaining ICAR variance parameters have inverse gamma prior distributions with shape and scale parameters of 0.5. 

Let $\bftheta_X$ denote the vector of all parameters from the covariate models. After adding this process level model to the hierarchical model in equation \eqref{eq:BHM}, the full posterior distribution is
\begin{equation}
    \begin{aligned}
          \pi\left(\bfZ_{1:T}, \bftheta_Z, \bfphi , \bftheta_X \mid \bfY_{1:T}, \bfX_{1:T} \right)  &\propto f\left(\bfY_{1:T} \mid \bfZ_{1:T}\right) \pi\left(\bfZ_{1:T} \mid \bftheta_Z, \bfX_{1:T}\right) \pi\left(\bftheta_Z \mid \bfphi\right) \pi\left(\bfphi\right) \\ 
        &\times f\left(\bfX_{1:T} \mid \bftheta_X \right) \pi\left(\bftheta_X\right) \\
        &\propto  \pi\left(\bfZ_{1:T}, \bftheta_Z, \bfphi \mid \bfY_{1:T}, \bfX_{1:T} \right) \times  \pi\left(\bftheta_X \mid \bfX_{1:T} \right),
    \end{aligned}
    \label{eq:postcov}
\end{equation}
which is the product of the posterior distribution given in equation \eqref{eq:fullpost} and the posterior distribution of the covariate model. So the full posterior in equation \eqref{eq:postcov} can be simulated by independent simulation from the two posterior distributions. 

The posterior distribution of the covariate model, $ \pi\left(\bftheta_X \mid \bfX_{1:T} \right)$, can efficiently be simulated using a two-stage MCMC algorithm. This follows as described in Section 3, so full details are redacted. However, the stage one model assumes the spatially-varying parameters are independent, which allows each location to be fit in parallel. The stage one samples are then used as proposed draws in a Metropolis-Hastings algorithm of the full model which assumes spatially-correlated parameters. Let $\left(\bfZ_{1:T}^{(m)}, \bftheta_Z^{(m)}, \bfphi^{(m)} , \bftheta_X^{(m)} \right)$ denote the $m$th sample from equation \eqref{eq:postcov}. Forecasts of drought are obtained by first simulating from equation \eqref{eq:covAR}, conditional on $\bftheta_X=\bftheta_X^{(m)}$, to obtain $\bfX_{T:1:T+\ell}^{(m)}$, and then simulating $\bfZ_{T+1:T+\ell}^{(m)}$ according to the model in equation \eqref{eq:Zmod}, conditional on $\bftheta_X^{(m)}$, $\bfZ_T^{(m)}$, and $\bfX_{T+1:T+\ell}^{(m)}$. The samples from the posterior predictive distribution for the ordinal drought levels are then generated by
$$Y_{i,T+\ell}^{(m)} = \sum_{j=0}^J j \cdot \textrm{I}(\alpha_j < Z_{i,t+\ell}^{(m)} \leq \alpha_{j+1}).$$ 

\begin{figure}[h]
    \centering
    \includegraphics[width=0.99\linewidth]{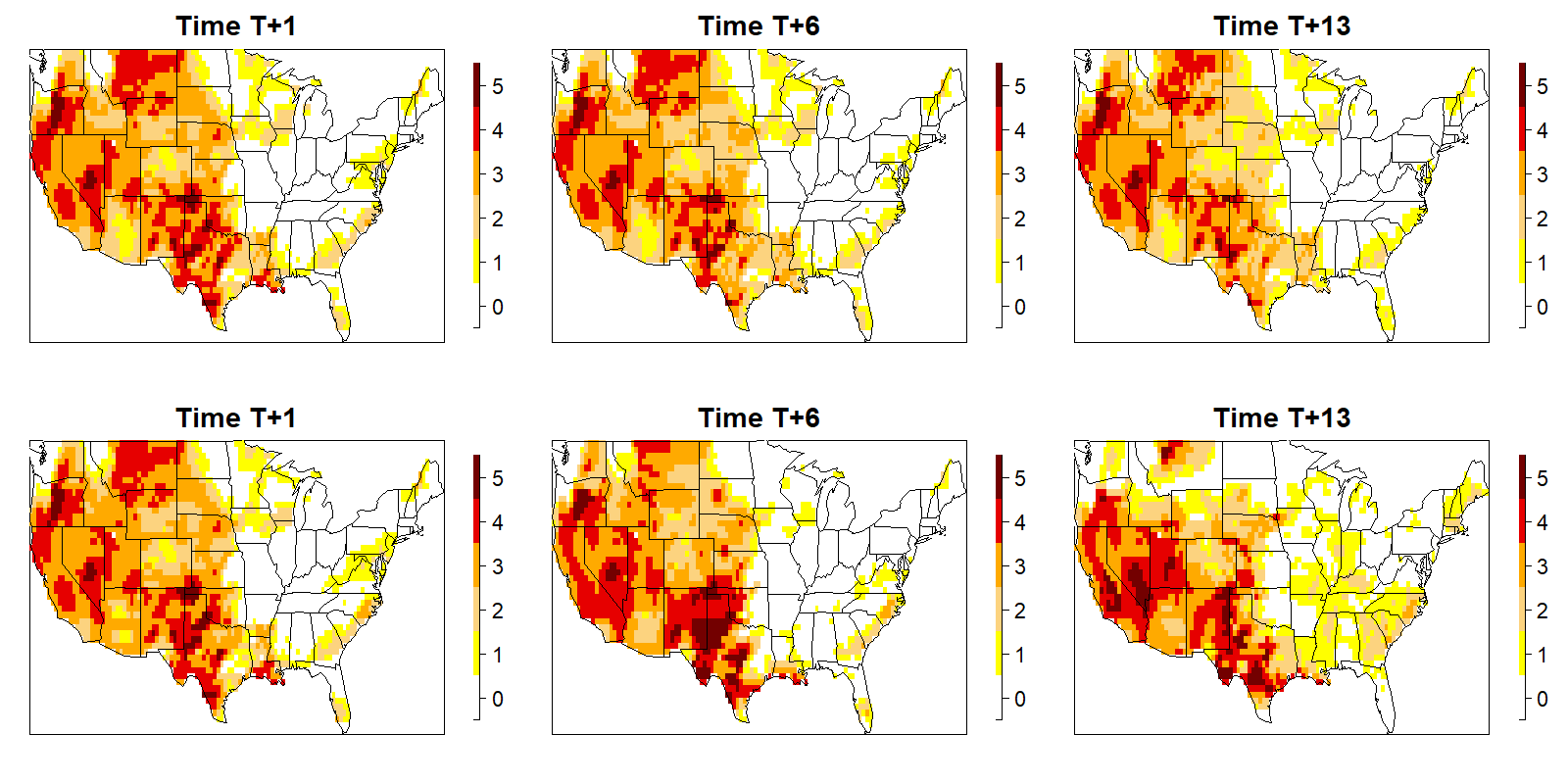}
    \caption{Predicted (top row, taken as the posterior predictive distribution median) and actual holdout (bottom row) drought levels for the next 1 (left), 6 (middle), and 13 (right) weeks.  Observe that the predictive accuracy decays as time increases.}
    \label{fig:postpred}
\end{figure}

\begin{figure}[h]
    \centering
    \includegraphics[width=0.99\linewidth]{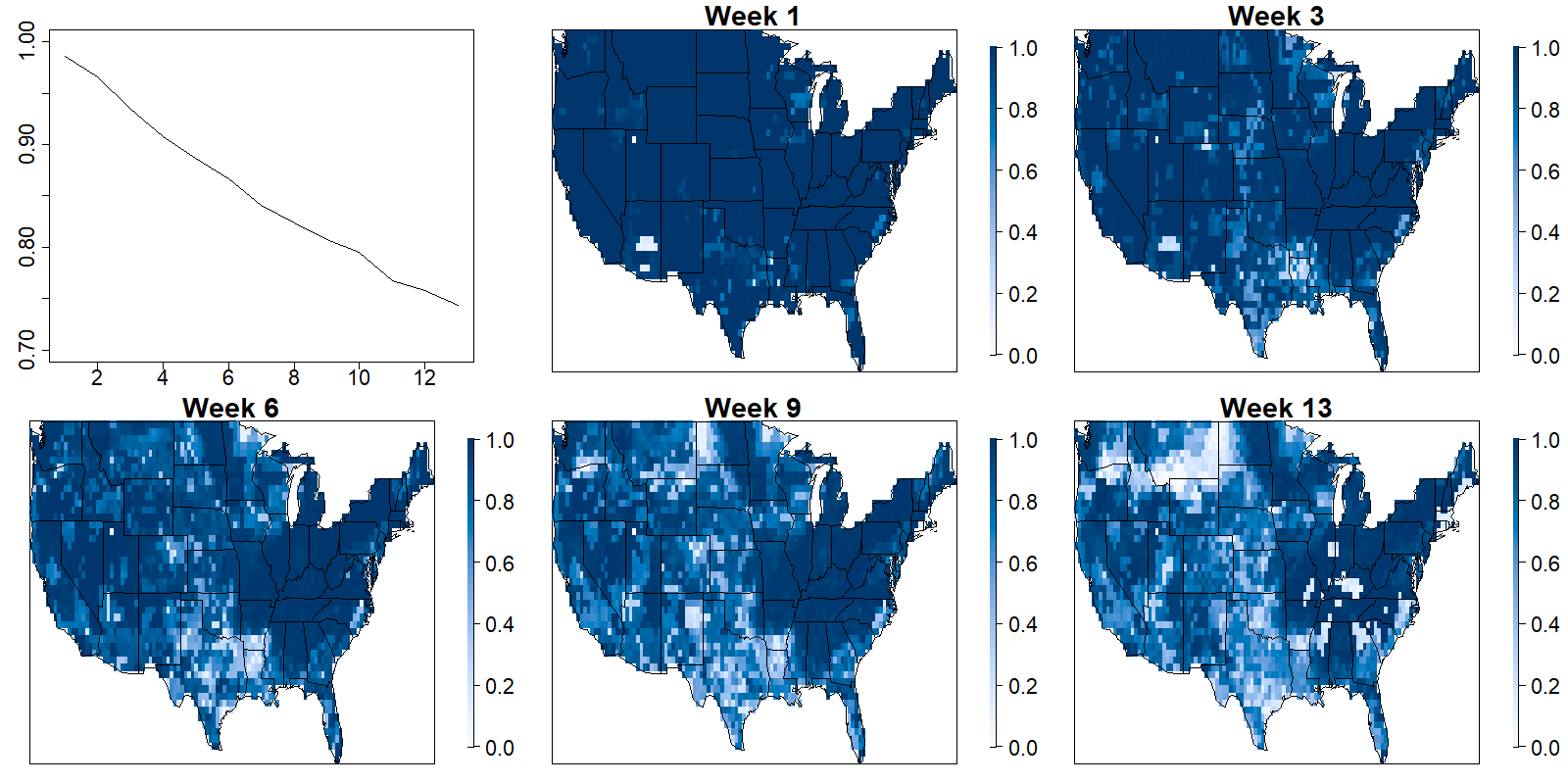}
    \caption{Maps of the posterior probability of being within one category of the actual drought level at 1, 3, 6, 9, 13 weeks. The top left plot is the probability averaged over all locations for each forecasted week. }
    \label{fig:decay}
\end{figure}

Figure \ref{fig:postpred} shows the median of the posterior predictive distribution of $Y_{i,T+\ell}$ at $T+1, T+6,$ and $T+13$ weeks.  The top row shows the posterior median as the point estimate for drought level by location.  The bottom row shows the actual drought level during that time period, as the holdout data are available for weeks $T+1$, ..., $T+13$.  Unsurprisingly, the correspondence at week one (left column) is very high, given the strong weekly predictability of drought as indicated in the autocorrelation parameters $\rho_i$ being very near 1.  As time advances through 13 weeks (middle and right columns), the correspondence decays.  To better illustrate and quantify this decay in predictive accuracy, we compute the posterior predictive probability that the drought level is within one of the true holdout drought level, by location and by week.  Figure \ref{fig:decay} shows these results.  The five figures labeled week 1, week 3, week 6, week 9, and week 13 all show what proportion of the posterior predictive distribution fell within one level of the true holdout.  These figures demonstrate a smooth decay from 95+\% of probability within one level (week 1) to around 75\% (week 13), when averaged over all grid cells.  The maps also show that there is spatial variability in these amounts.

\section{Discussion}

Our model and application described the specific case of an ordinal response variable $Y_{i,t}$, based on a latent spatio-temporal process $Z_{i,t}$ with location-specific parameter $\bftheta_{Z,i} = (\bfbeta_i, \rho_i, \sigma_i)$.  The location-specific parameters are modeled with  spatial priors, which accounts for spatial dependence across locations, but in general comes at a drastic increase in computational cost of model fitting. We proposed a two-stage MCMC algorithm that yields computationally efficient fitting of the Bayesian spatio-temporal model. Stage one fits the simplified model that assumes independence across all locations, in parallel. The samples from the stage one posterior are then used as proposed values for a Metropolis-Hastings algorithm of the full posterior distribution in stage two. 

While the motivation of our development was for spatio-temporal ordinal data, the two-stage MCMC approach extends far beyond this example to a broad class of spatio-temporal applications, which was illustrated by our second use of this algorithm in fitting the multivariate spatio-temporal model for the covariates in Section 5. Specifically, what is needed to implement this approach is a model of the form $Z_{i,t} \mid \bftheta_i$ with location-specific parameter $\bftheta_i$. In the case of a continuous response variable, $Z_{i,t}$ could be observed data. But as was illustrated in our application, this approach also holds in non-normally distributed settings when the observed data relate to an underlying continuous process. For example, this approach can be applied for spatio-temporal generalized linear models where $Z_{i,t}$ represents the link function for a particular data model.



Although our proposed approach is broadly applicable to a large class of spatio-temporal models, there are several limitations that we discuss here. The first is that the second stage involves a Metropolis-Hastings update, in which the entire parameter $(\bfZ_{i,1:T}, \bftheta_i)$ is updated jointly in a single step.  The acceptance probability in equation \eqref{eq:R} depends on $\bftheta_i$, and depending on the dimension and model specification of this vector, the acceptance probability $R$ could be vanishingly small, in which case the MCMC could mix very poorly.  In our application, we considered a joint update whose parameter vector had six elements --- four for $\bfbeta_i$, and one for each of $\rho_i$ and $\sigma_i^2$ --- and the acceptance probability $R$ depends on the five parameters that had spatial priors in the full model. If we had considered an application with far more covariates and a larger dimension $\bfbeta_i$, then the second-stage MCMC would have updated much less frequently.  Thus, in its current formation, the two-stage methodology does not scale well with the dimension of the location-specific parameter vector.

A second limitation concerns the relationship between the quality of the stage one posterior and the stage two posterior.  The second stage uses the posterior from stage one as the proposal density in a Metropolis-Hastings algorithm.  It should therefore be clear that stage-two posterior draws are simply a re-weighting of stage one posterior draws.  Accordingly, the two-stage methodology will only work if stage one yields a ``good'' approximation to the posterior in the sense that it explores the full parameter space. A stage one posterior which fails to explore some regions of the parameter space would not permit the second stage to explore that region either. Accordingly, users must be cognizant that the stage one model posterior distribution has support that contains the support of the full model and that both the stage one and stage two MCMC mix well and converge to their respective stationary distributions. 

A third consideration is that this methodology requires multiple replicates or data from multiple time periods for each location. For spatial data (where the number of time periods is $T=1$), the methodology as proposed would not work with a single realization of the process $\bfZ_i$. This is because our stage one model assumed independence across locations, and that model would not be identifiable with just a single observation. Future work aims to modify this methodology to enable similar computational benefits in high-dimensional space-only problems. 


Despite these considerations, the methodology we developed extends the reach of fitting spatio-temporal models to larger datasets by substantially reducing the computational cost.  When compared to a single-stage MCMC model fit, our two-stage model fit permits parallel computing in stage one through independent models, and is followed by a computationally efficient resampling algorithm in stage two which brings back the spatial dependence in the posterior distribution through spatial priors. The resulting posterior distribution allows for posterior inference at a fraction of the single-stage computational cost.  Our method  also allows for spatio-temporal models to be fit to substantially larger datasets at a reasonable computational cost.  Finally, it lays the groundwork for fitting Bayesian spatio-temporal models to a set of data types which include binary, count, and quantitative responses in a spatio-temporal setting.


\section{Data and Code Availability}
All data are freely available at Data Dryad \url{https://datadryad.org/stash/dataset/doi:10.5061/dryad.g1jwstqw7#citations} with code available at \url{https://github.com/heplersa/USDMdata}.  The data are described in extensive detail in \citet{erhardt2024homogenized}.  Code from this paper is available in \url{https://github.com/heplersa/DroughtModelTwoStageMCMC}.

\section{Conflict of Interest}
The authors declare no conflicts of interest.

\section{Funding Statement}
The authors acknowledge support from NSF award \#2151881.

\bibliographystyle{jasa}
\bibliography{droughtref} 

\newpage

\newpage
\appendix
\counterwithin{figure}{section}
\section{Appendix/Supplemental}

 \subsection{Illustration of Second Stage Impact}

\begin{figure}[H]
    \centering
    \includegraphics[width=0.95\linewidth]{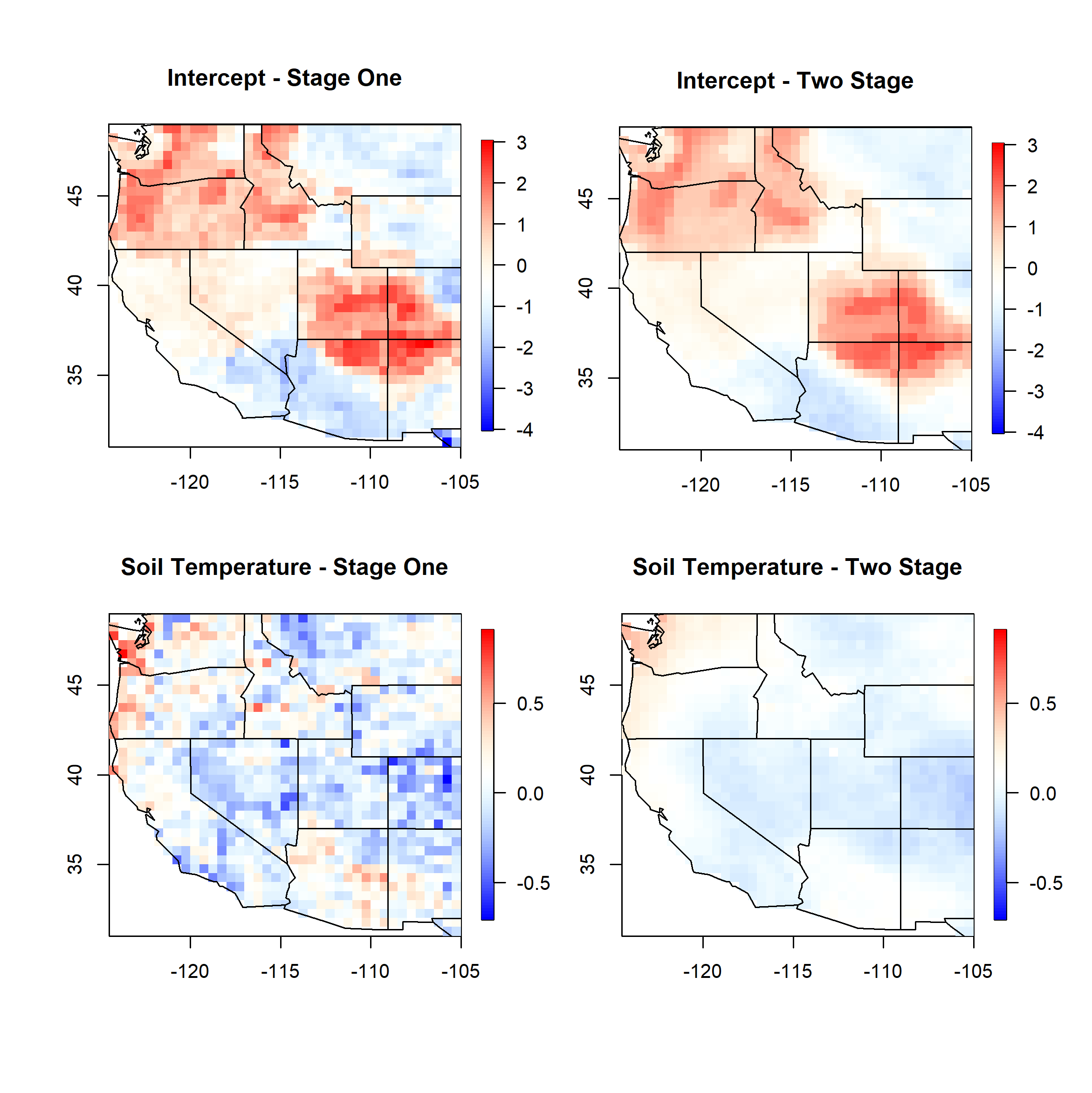}
    \caption{Left column: posterior means for the intercept and soil moisture parameters after the (independent, parallel) stage one model fit.  Right column: updated posterior means after the stage-two resampling.  Observe the spatial smoothing and dependence captured during the stage two model fit.}
    \label{fig:stage2impact}
\end{figure}

We use the reduced data set to demonstrate the distinction between the first stage and second stage in the two-stage method. Figure \ref{fig:stage2impact} shows posterior means for two of the parameters, $\beta_0$ and $\beta_{soilm}$ (the others parameter posterior mean results are substantially similar and not shown due to space limitations).  The left column shows the posterior mean after the first stage of the two-stage algorithm, which fits site-specific models at each location, independently and in parallel.  Therefore, only information from location $i$ is utilized when sampling from the posterior for $\bfbeta_i$.  The right column shows the posterior mean obtained after the second stage, which resamples from the first stage using a Metropolis-Hastings update whose probability of acceptance is shown in equation \eqref{eq:R}. Observe that the right panel demonstrates more spatial smoothness because the full model assumes spatial random effects for this parameters.

\subsection{Additional Comparisons between Single Stage and Two Stage Methods for the Western United States}

\begin{figure}
    \centering
    \subfigure[]{\includegraphics[width=.32\textwidth]{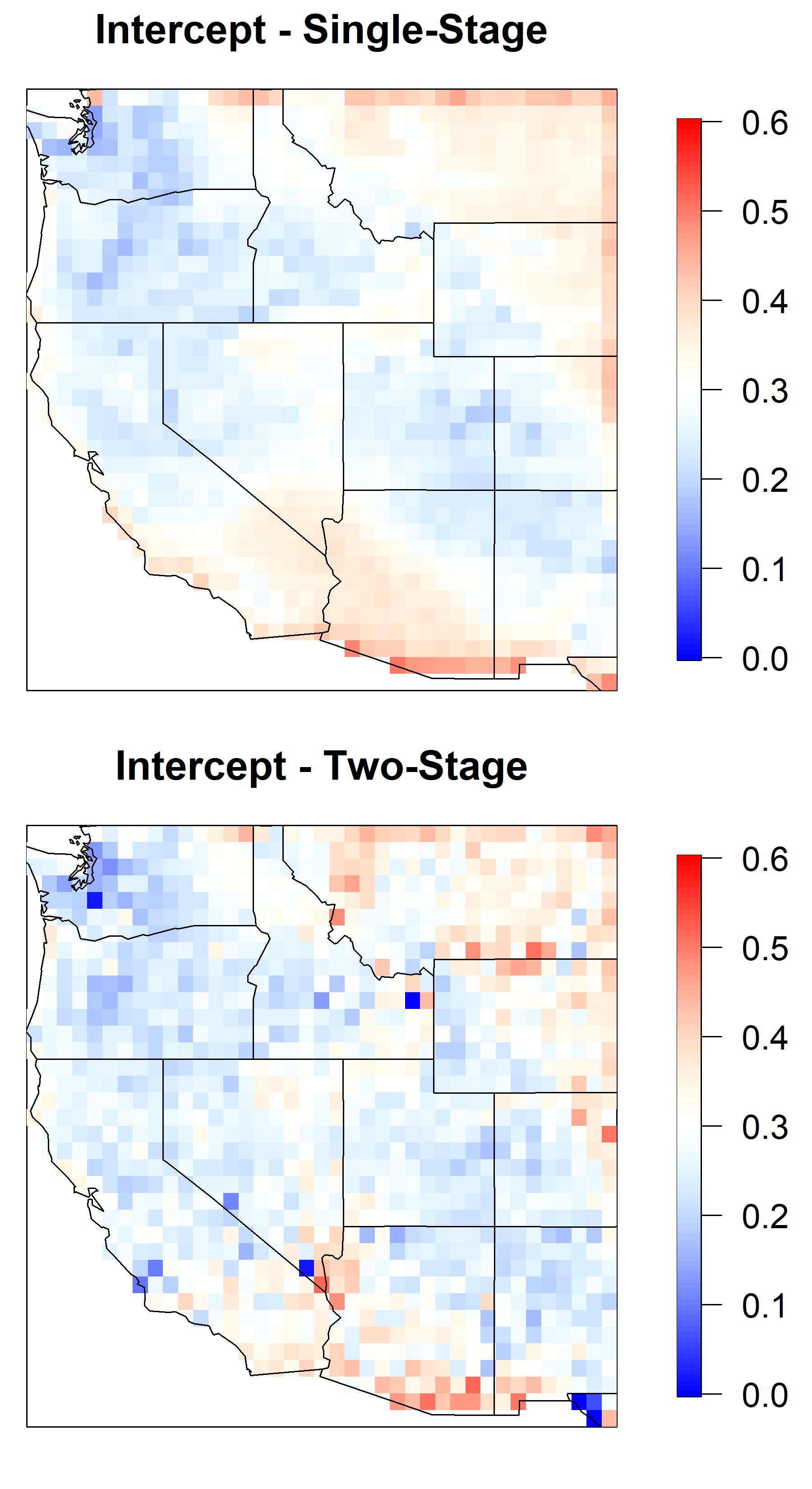}}
    \subfigure[]{\includegraphics[width=.32\textwidth]{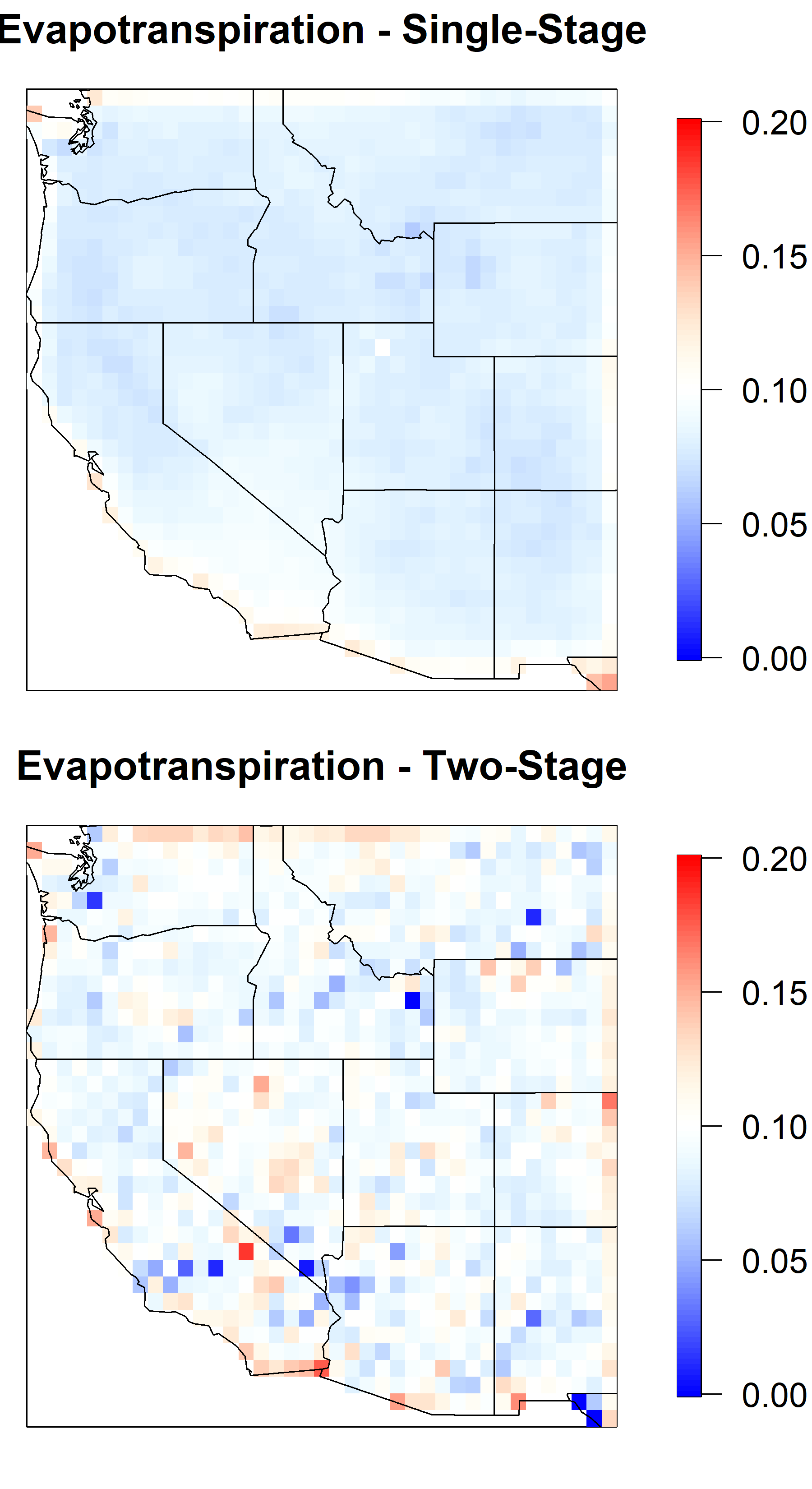}}
    \subfigure[]{\includegraphics[width=.32\textwidth]{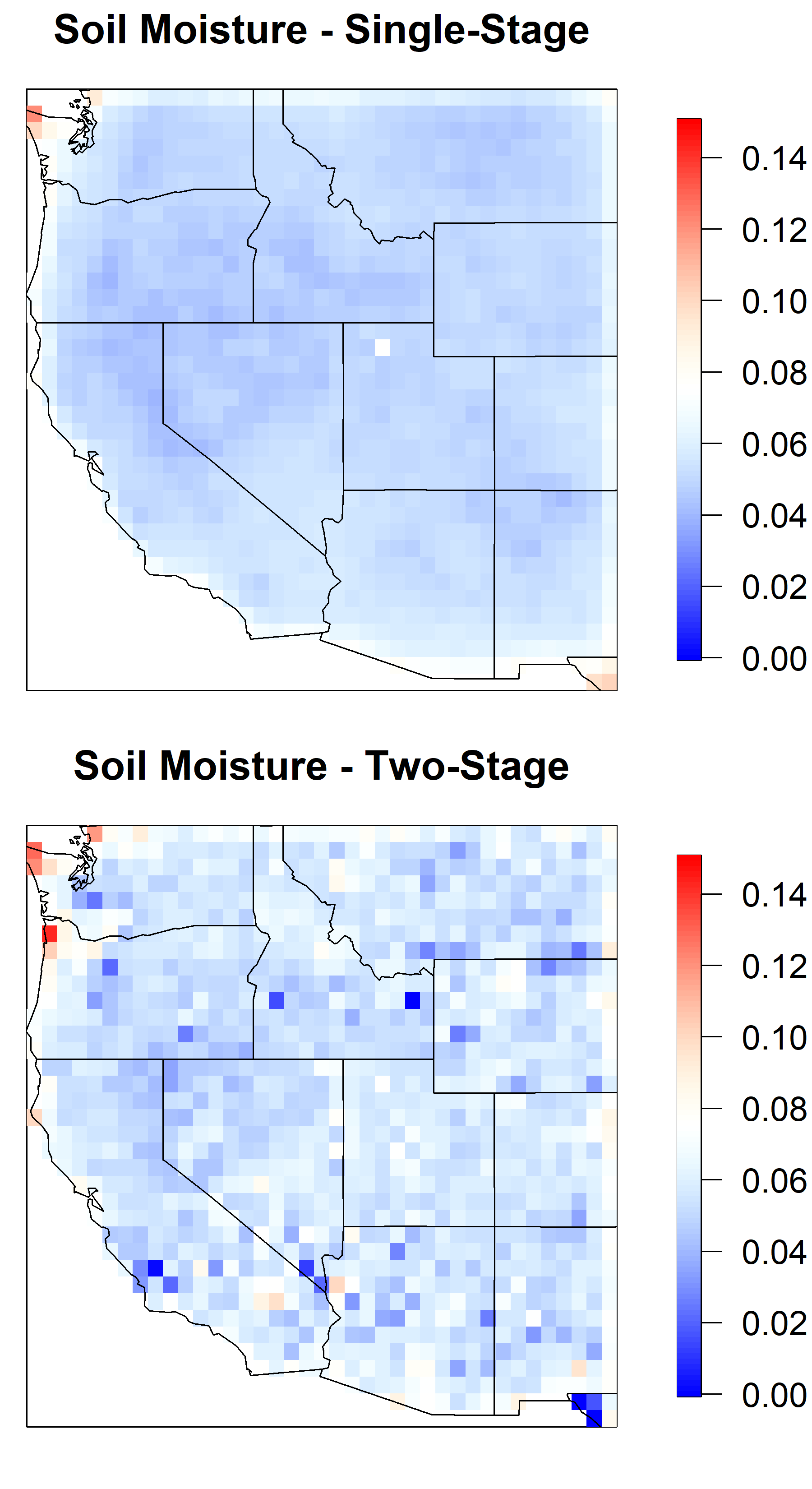}}
    \subfigure[]{\includegraphics[width=.32\textwidth]{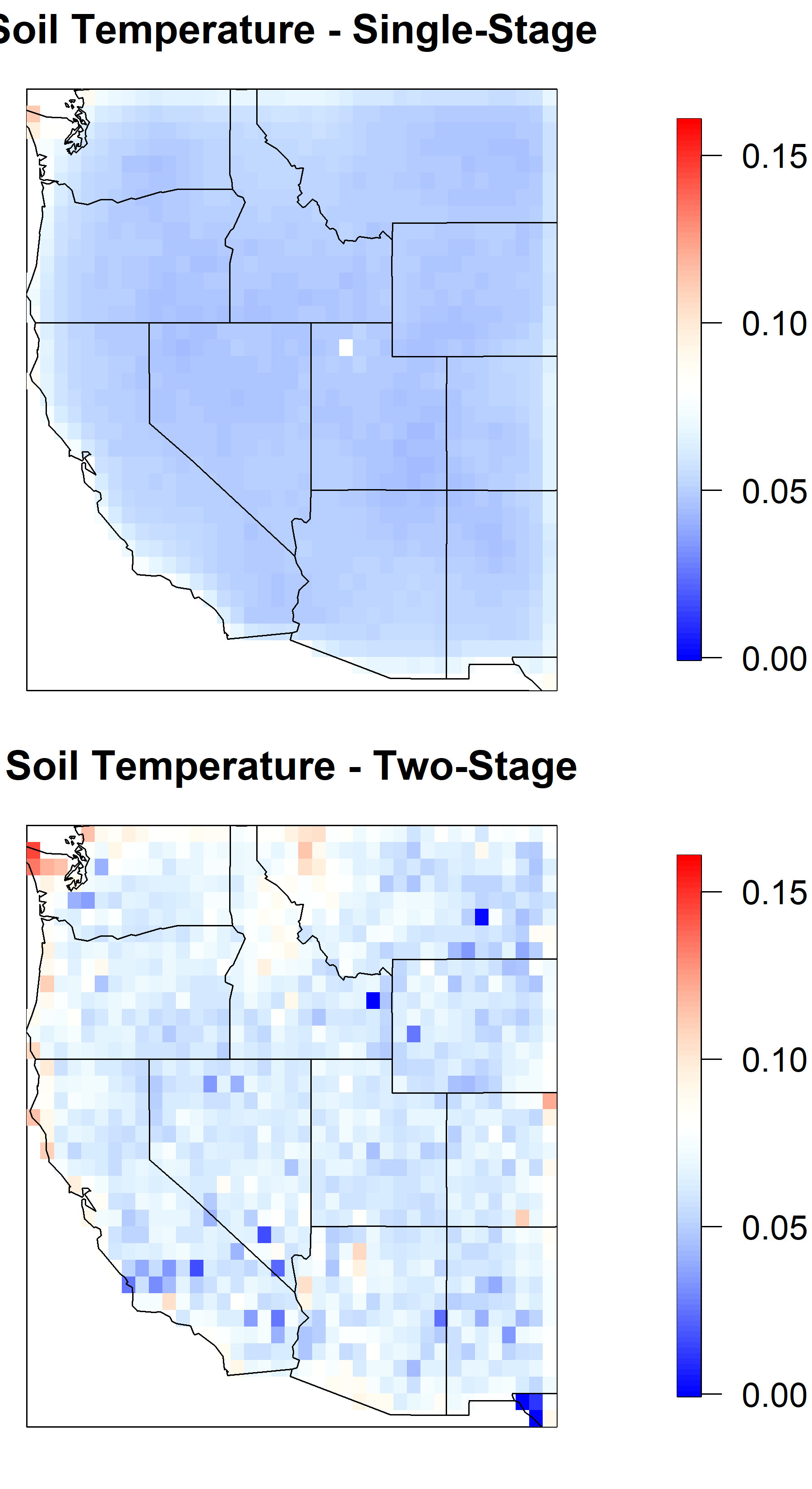}}
    \subfigure[]{\includegraphics[width=.32\textwidth]{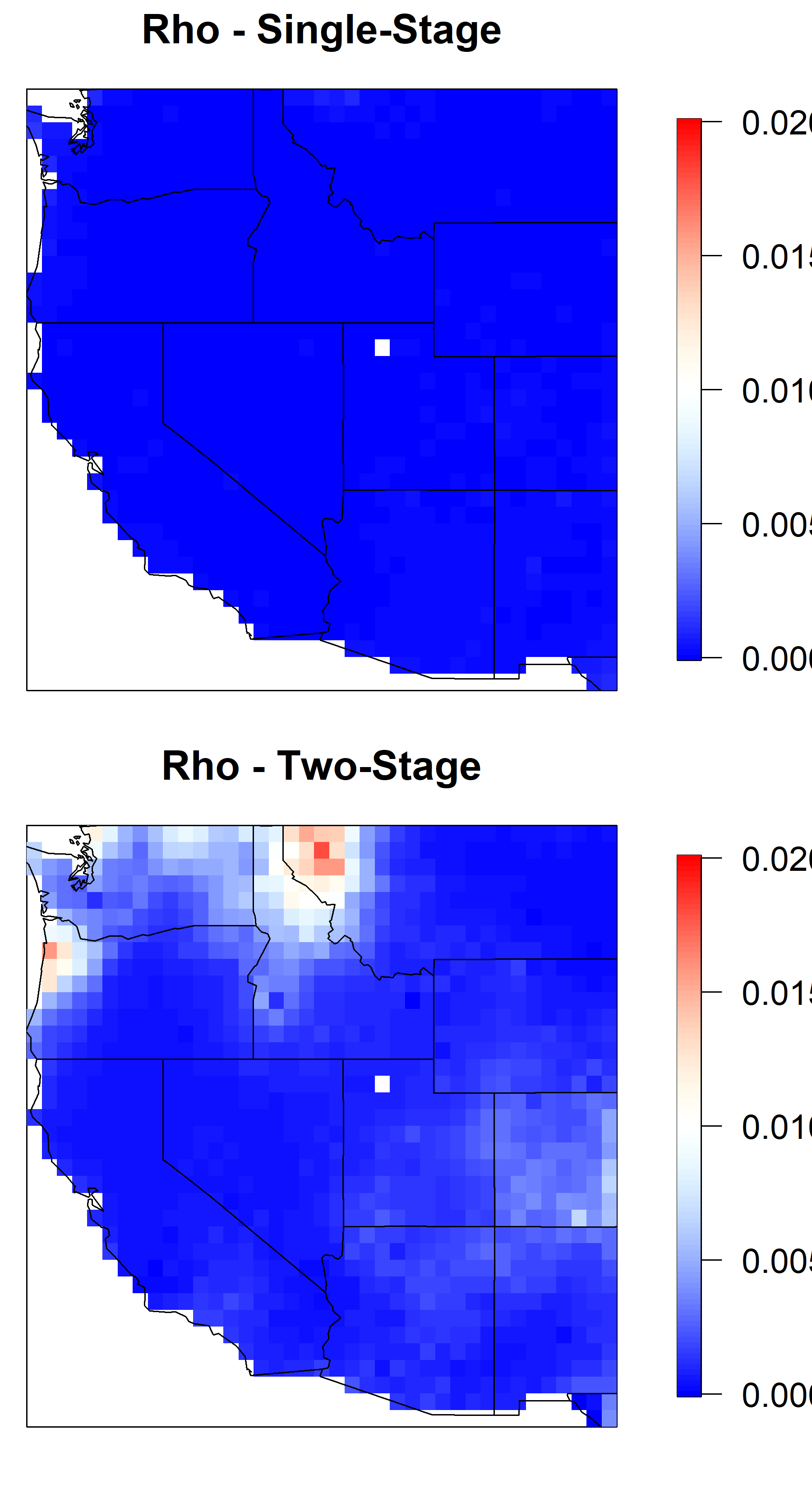}}
    \subfigure[]{\includegraphics[width=.32\textwidth]{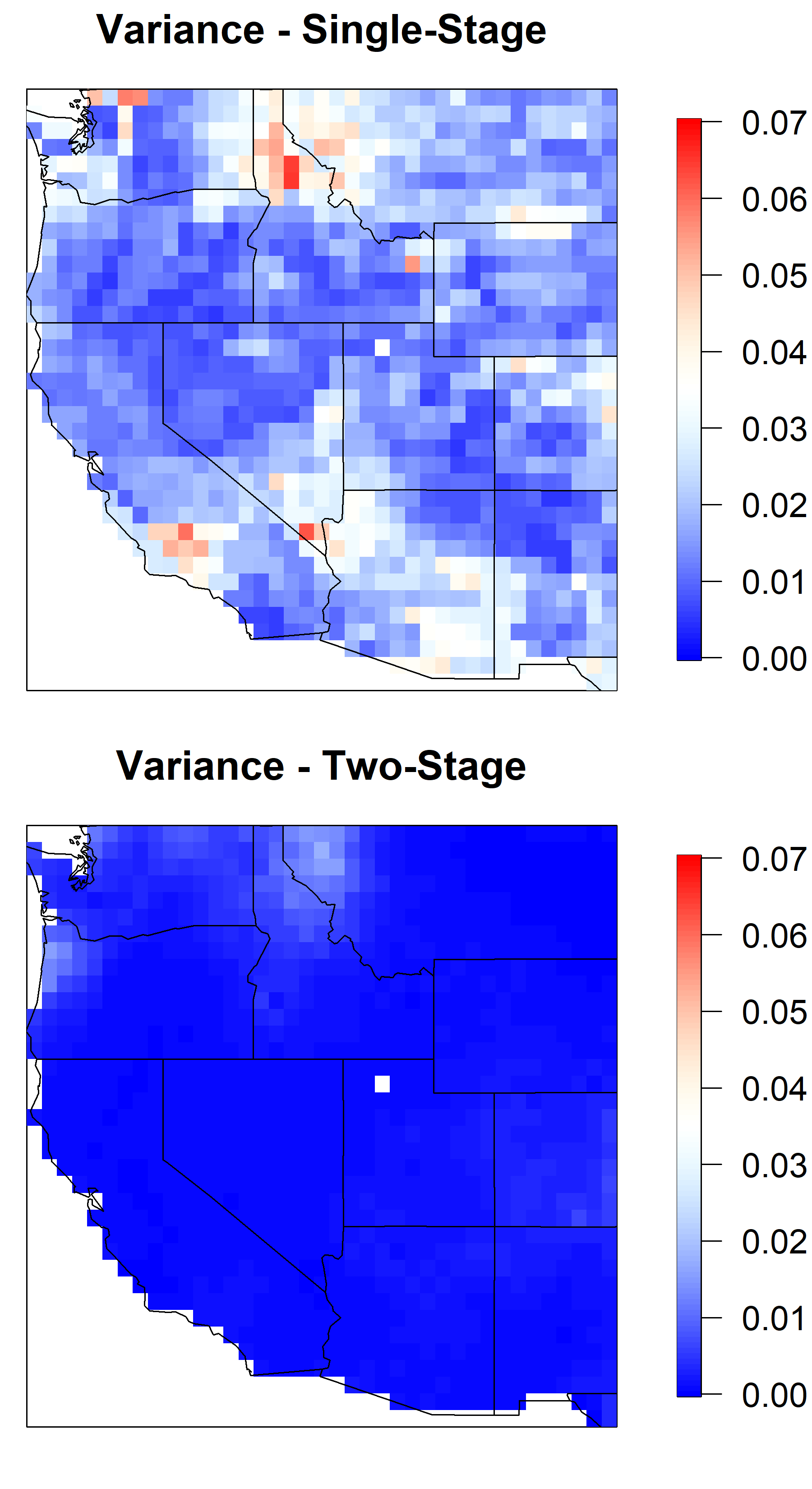}}
    \caption{Posterior standard deviations of the regression coefficients (panels (a)-(d)), temporal autocorrelation parameter (panel (e)), and variance parameter (panel (f)) for the single-stage MCMC (top row) and two-stage MCMC (bottom row). }
    \label{fig:singlevstwoSD}
\end{figure}

\begin{figure}
    \centering
    \subfigure[Posterior SD of intercept $\beta_0$]
    {\includegraphics[trim={1cm 1cm 0 0cm},clip, page=1,width=0.49\textwidth, height=5.5cm]{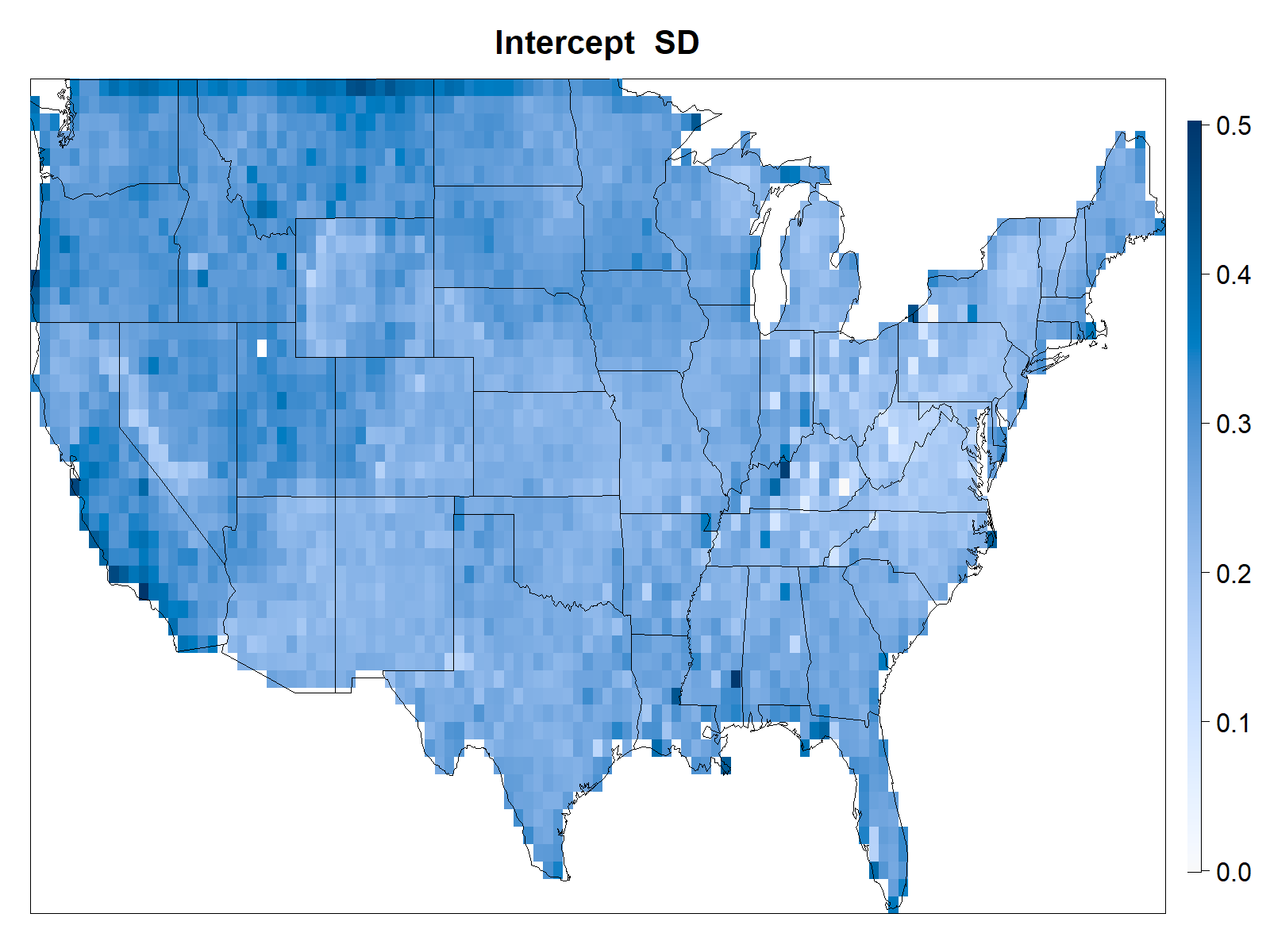}}  
    \subfigure[Posterior SD of evapotranspiration $\beta_1$]{\includegraphics[trim={1cm 1cm 0 0cm},clip,page=3,width=0.49\textwidth, height=5.5cm]{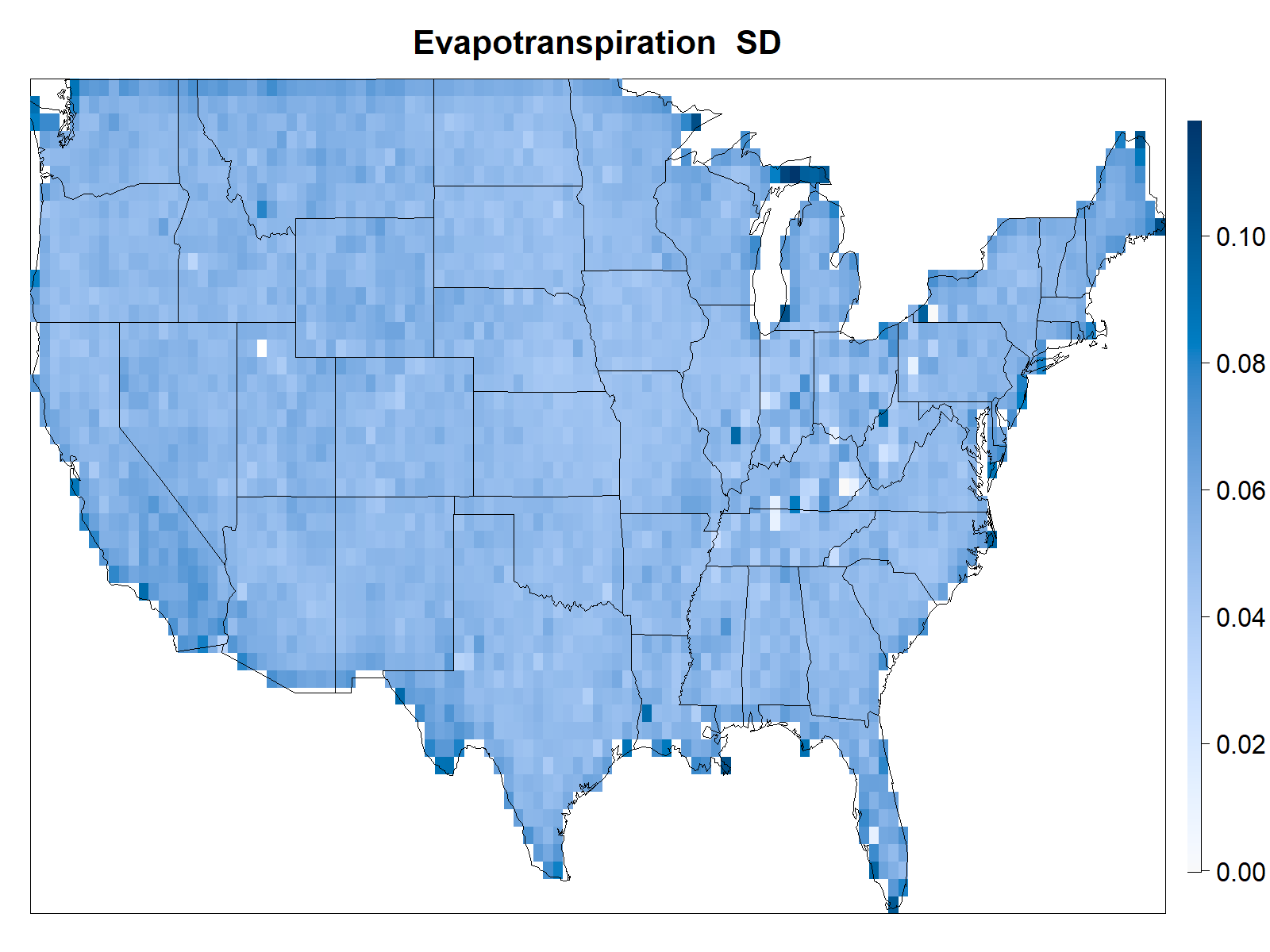}} 
    \subfigure[Posterior SD of soil moisture $\beta_2$]{\includegraphics[trim={1cm 1cm 0 0cm},clip,page=5,width=0.49\textwidth, height=5.5cm]{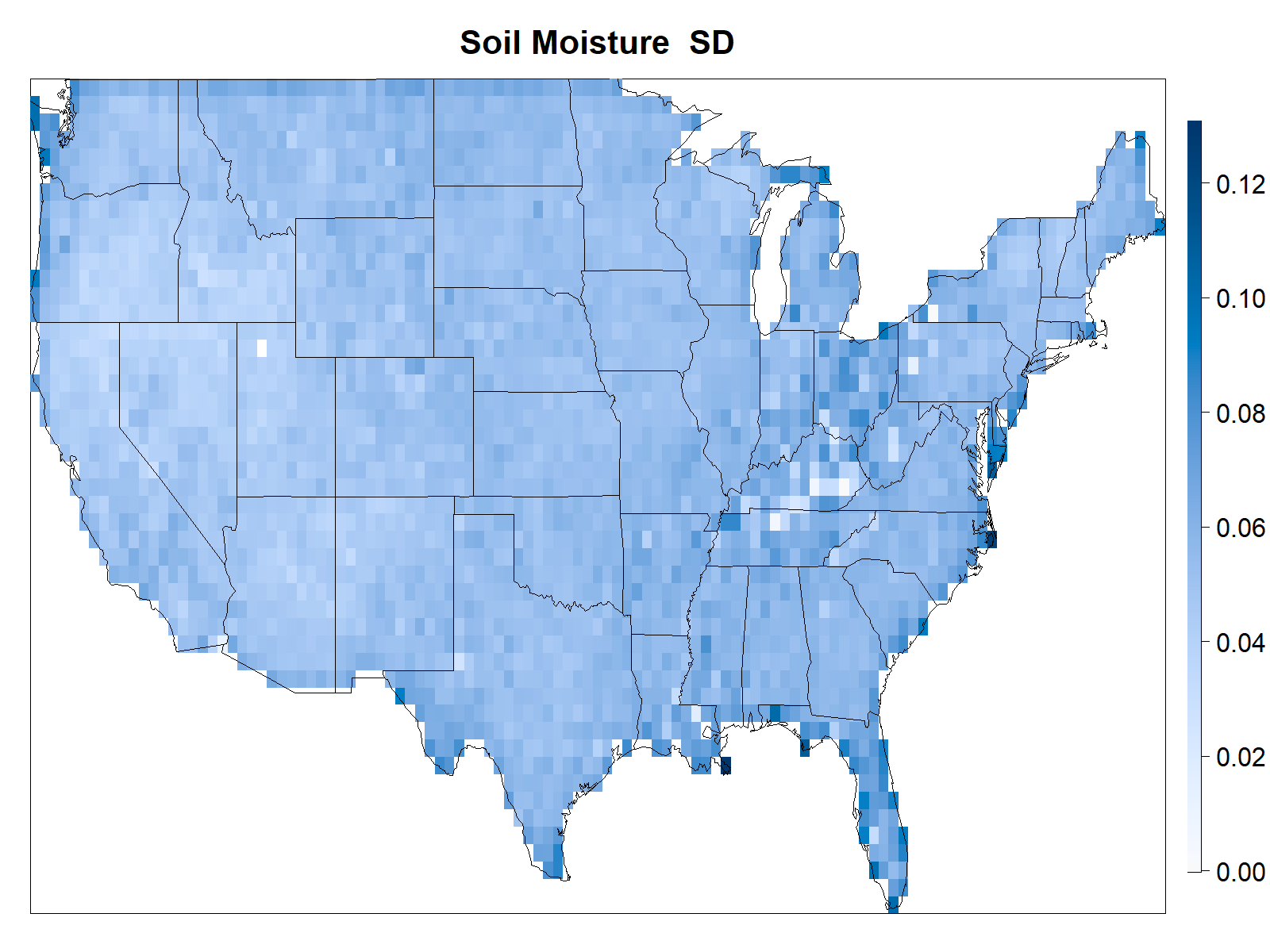}}
    \subfigure[Posterior SD of soil temperature $\beta_3$]{\includegraphics[trim={1cm 1cm 0 0cm},clip,page=7,width=0.49\textwidth, height=5.5cm]{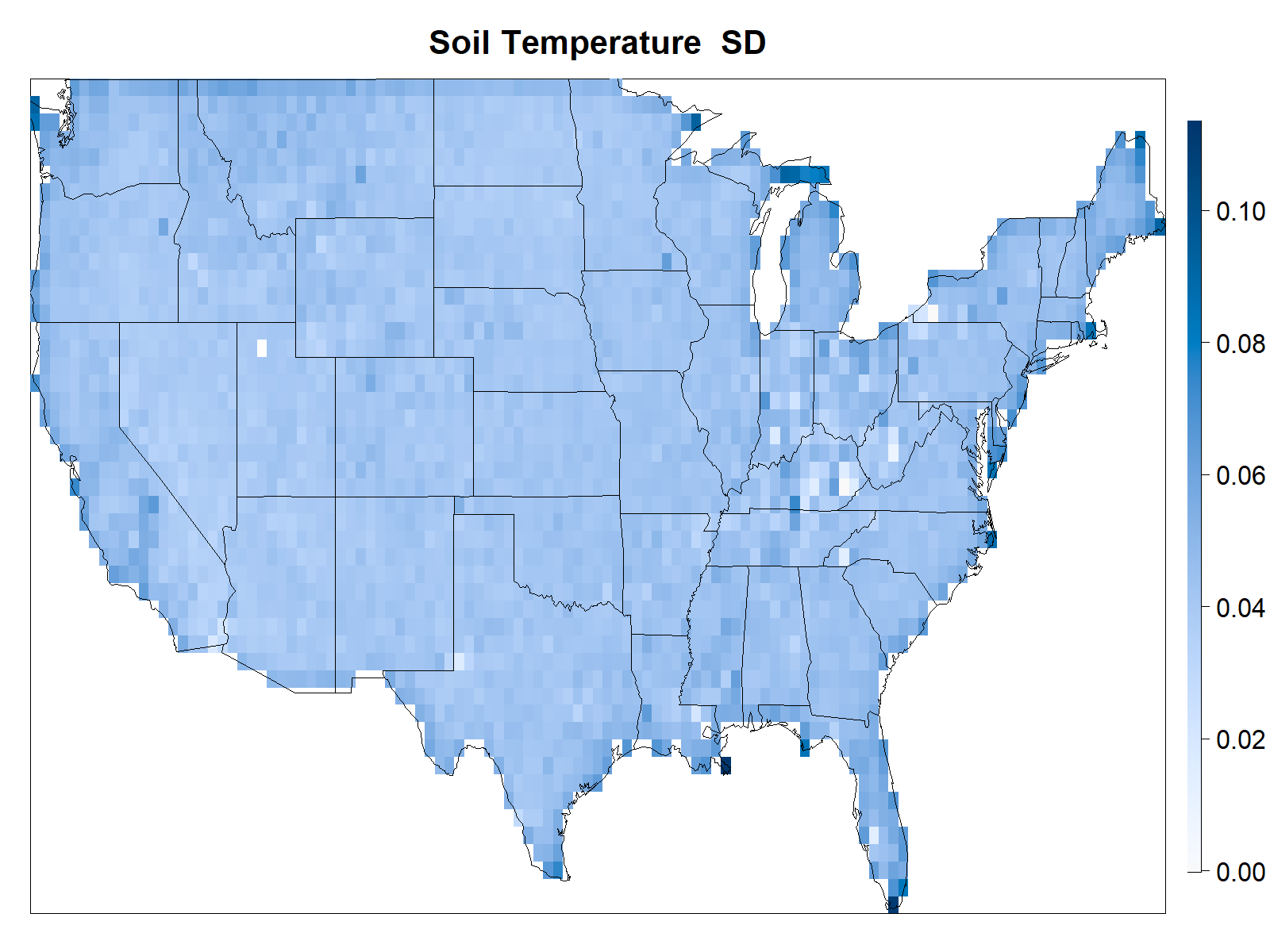}}
    \subfigure[Posterior SD of variance $\sigma^2$]{\includegraphics[trim={1cm 1cm 0 0cm},clip,page=11,width=0.49\textwidth, height=5.5cm]{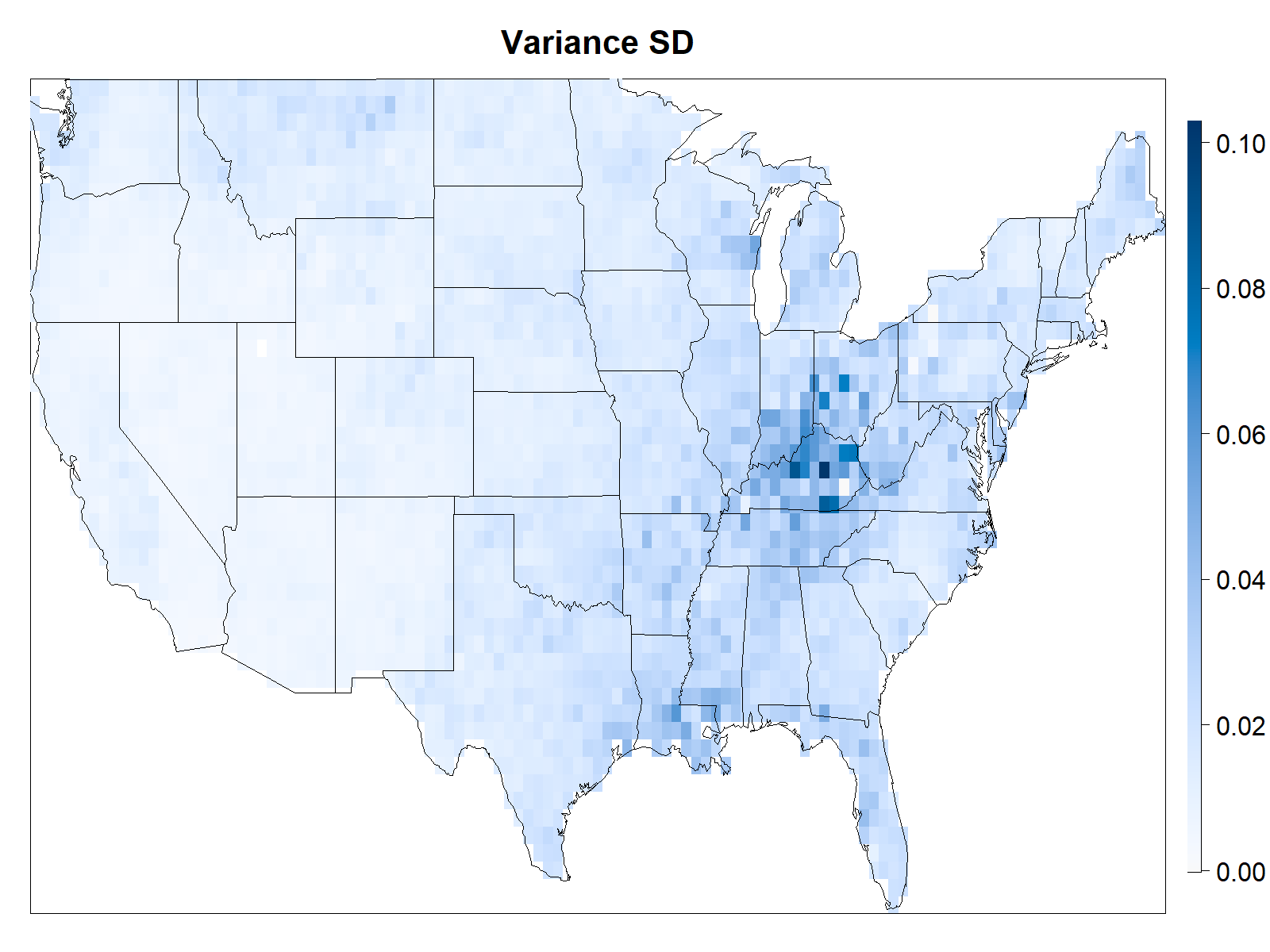}} 
    \subfigure[Posterior SD of autoregression  
 $\rho$]{\includegraphics[trim={1cm 1cm 0 0cm},clip,page=9,width=0.49\textwidth, height=5.5cm]{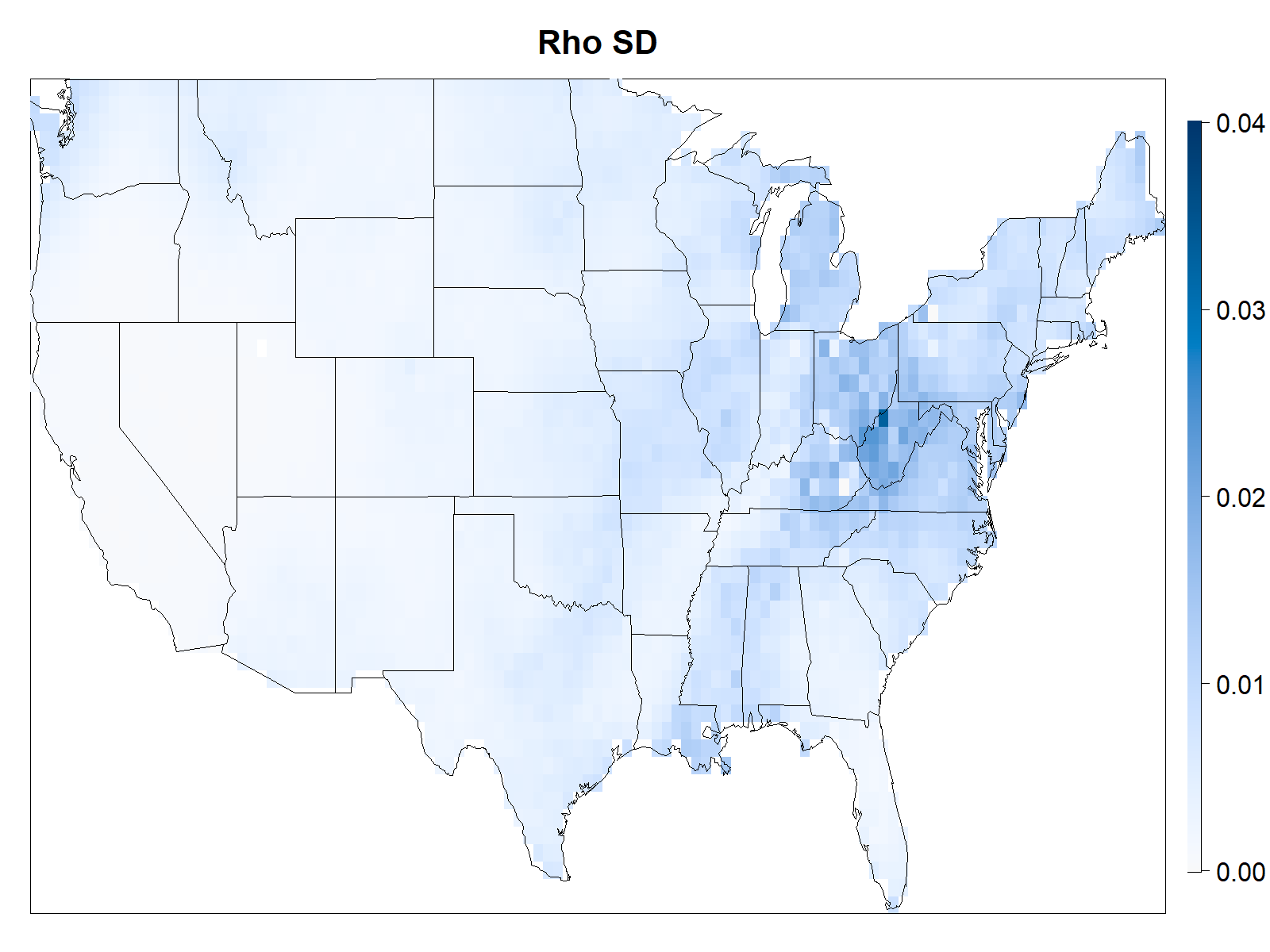}} 

    \caption{Posterior standard deviations for the regression coefficients taken from the two-stage model, implemented for the US.  Parameters shown include: (a) intercept $\beta_0$; (b) effect of evapotranspiration $\beta_1$; (c) effect of soil moisture $\beta_{2}$; (d) effect of soil temperature $\beta_{3}$; (e) variance $\sigma^2$; and (f) autoregression $\rho$.}
    \label{fig:betasd}
\end{figure}

Figure \ref{fig:singlevstwoSD} shows posterior standard deviations for the Western United States, for each of the parameters highlighted in figure \ref{fig:singlevstwo}, for both the single-stage method (top panels) and two-stage method (bottom panels).  Figure \ref{fig:betasd} shows posterior standard deviations for main parameters in the full United States application of the two-stage methodology.




\newpage
\subsection{Example of the Covariate Model Fit}
Here we show results from the covariate model fit for one grid cell, R87, to demonstrate the model fits for the model described in equation \eqref{eq:ts}.  This grid cell was randomly selected as a representative grid cell, and model fits from other grid cells are substantially similar.

\begin{figure}[!h]
{\includegraphics[width=6.5in]{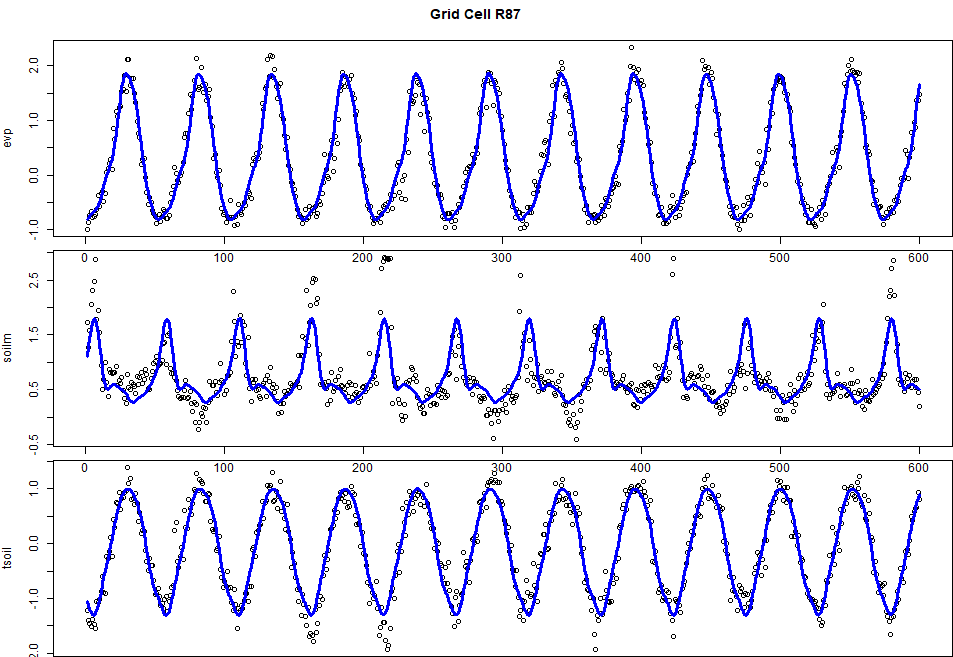}}
{\caption{An example of the time series model fit for the covariate mean functions shown in equation \eqref{eq:ts}.  The three plots show mean function fits for each of the three covariates, all for grid cell R87.  Data and fits from other grid cells are substantially similar.}\label{fig:S1}}
\end{figure}

\begin{figure}[!h]
{\includegraphics[width=6.5in, height=4in]{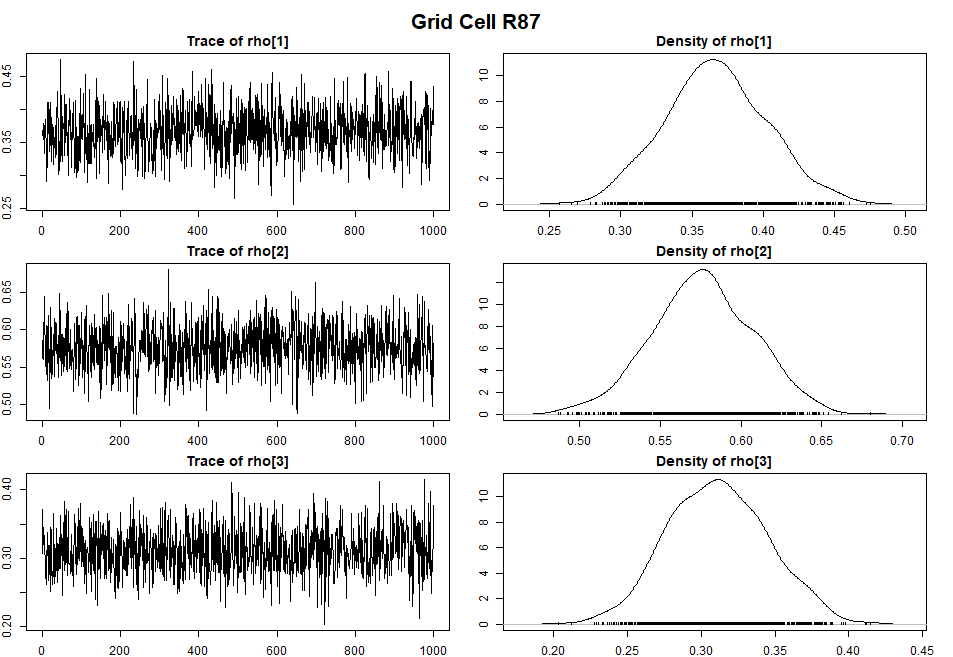}}
{\caption{Trace plots for the three autoregresive parameters $\delta_j, j=1, 2, 3$ for grid cell R87.}\label{fig:S2}}
\end{figure}

\begin{figure}[!h]
{\includegraphics[width=6.5in, height = 4in]{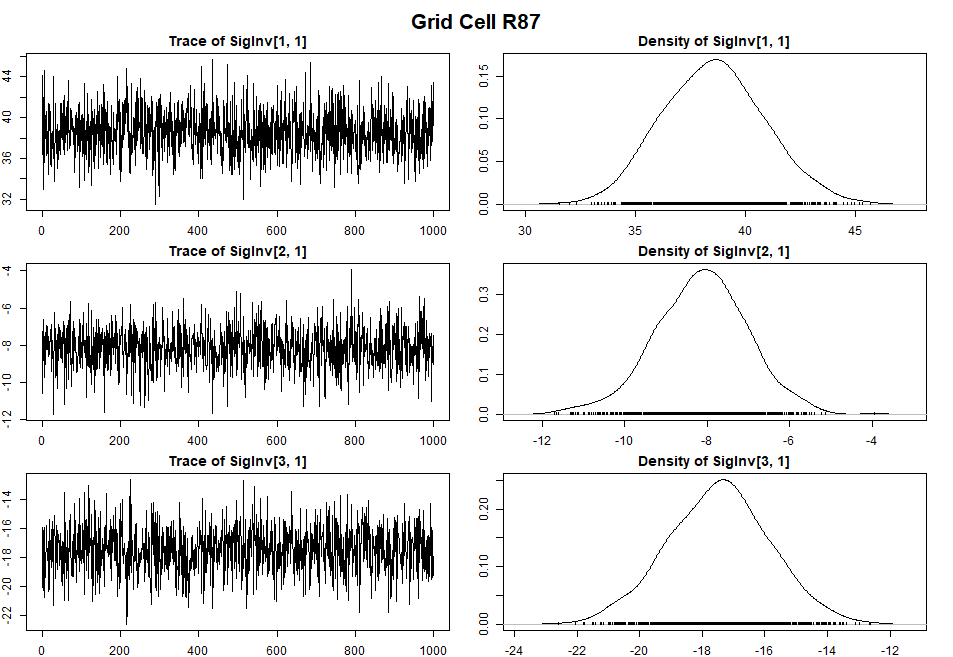}}
{\includegraphics[width=6.5in, height = 4in]{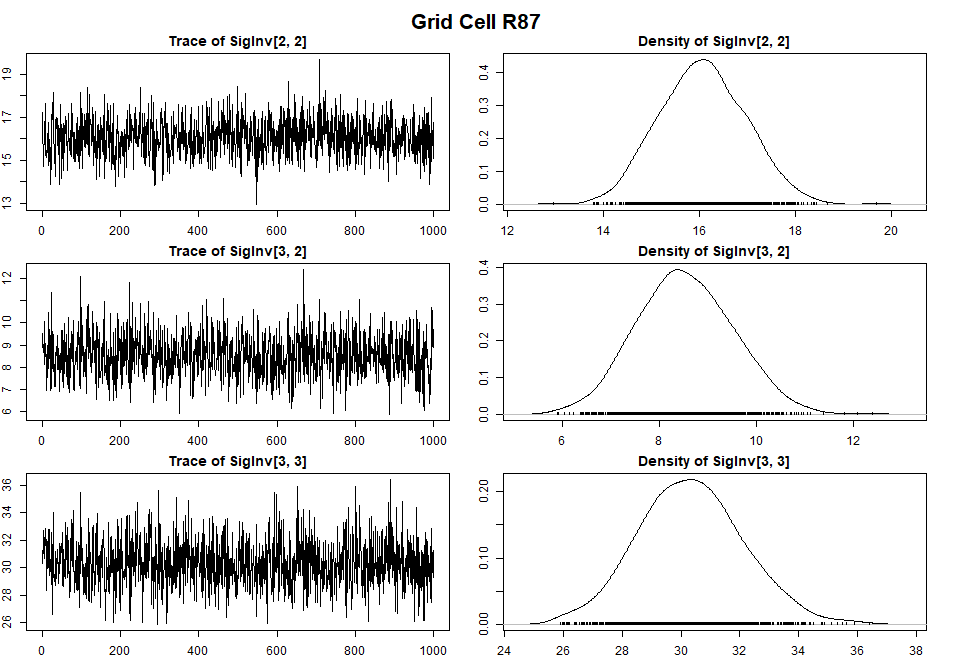}}
{\caption{Trace plots for the six unique parameters of the covariance matrix parameters in $\bf{\Sigma}$ for grid cell R87.}\label{fig:S3}}
\end{figure}

\begin{figure}[!h]
\begin{center}
{\includegraphics[width=3.5in]{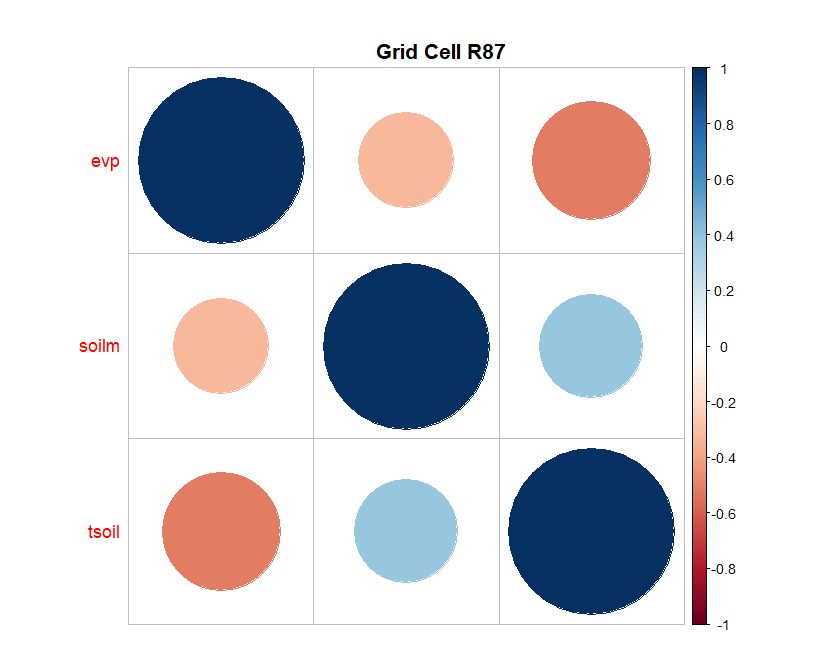}}
{\caption{Correlation plot of the posterior means of the covariance matrix parameters in $\bf{\Sigma}$.  Observe the mixture of positive and negative dependence across the three covariates.}\label{fig:S5}}
\end{center}
\end{figure}

Figure \ref{fig:S1} shows the time series of the three covariates, along with the fitted periodic function described in equation \eqref{eq:ts}.  Trace plots and posterior distributions of the autoregressive parameters $\delta_j$ and covariance elements of $\Sigma$ are shown in figures \ref{fig:S2} and \ref{fig:S3}.  A correlation plot of the posterior means of $\Sigma$ is shown in figure \ref{fig:S5}, which demonstrates that some pairs of environmental covariates display positive dependence, while others display negative dependence.  

\newpage
\clearpage
\subsection{Example of Covariate Forecasts and Prediction Error}

\begin{figure}[!h]
{\includegraphics[width=6.5in]{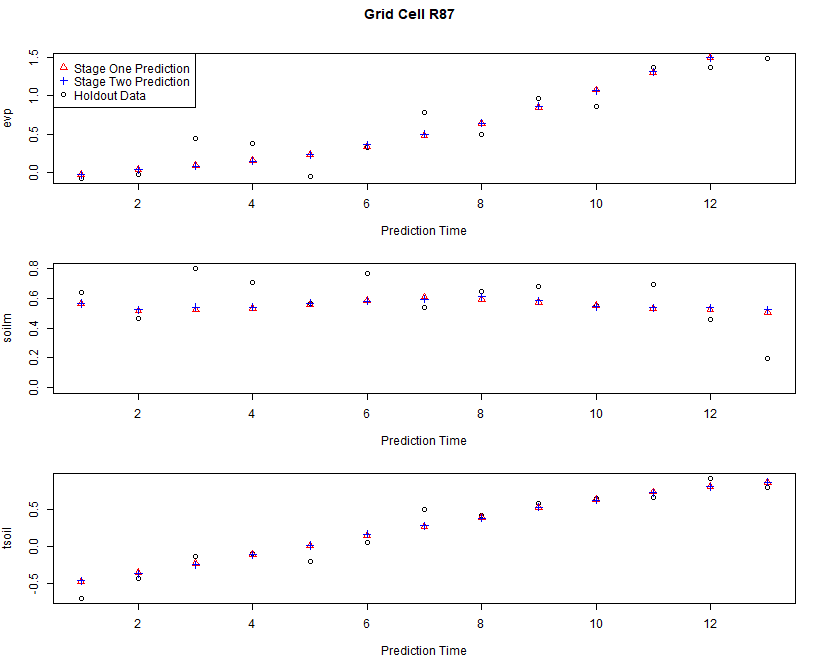}}
{\caption{The holdout observed data over the thirteen week holdout period, along with model's posterior mean predictions after the first stage and second stage (red triangles and blue plusses, respectively).}\label{fig:S6}}
\end{figure}

\begin{figure}[!h]
{\includegraphics[width=6in]{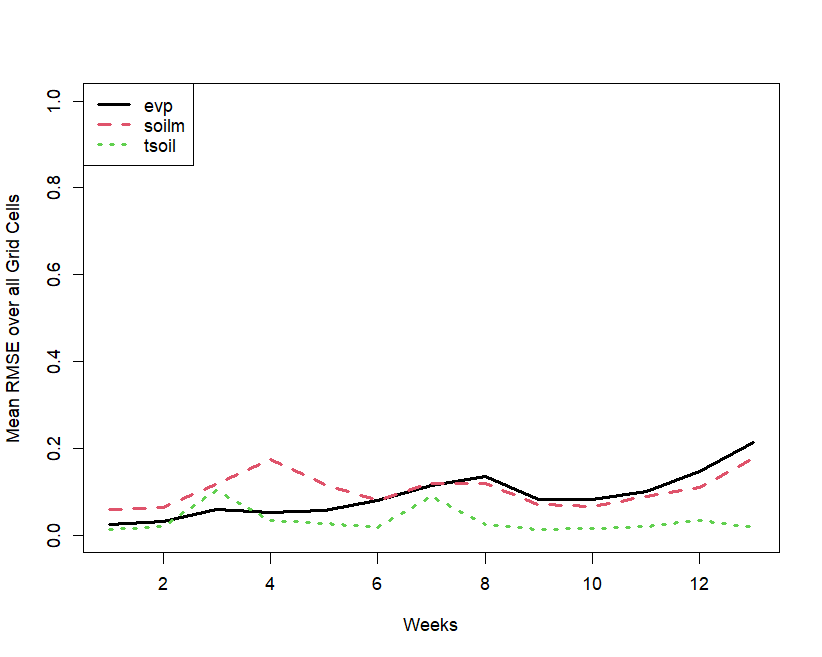}}
{\caption{Empirical root mean square errors between the holdout observed data and covariate model predictions, averaged across all locations.}\label{fig:S8}}
\end{figure}

Section 5.2 describes the need for posterior predictions of the covariates $\bfX_{T+1:T+\ell}$ for up to $\ell$ time periods into the future.  These are required to simulate from the posterior predictive distribution $\bfY_{T+1:T+\ell} \mid \bfY_{1:T}$, which is ultimately used to make forecasts of the ordinal response variable with uncertainty.  Figure \ref{fig:S6} shows the holdout observed data over a 13 week prediction period, along with posterior mean predictions taken after the first stage and second stage of the two-stage model, all from the same grid cell R87 whose results were shown above.  From this, the root mean square error (RMSE) of stage-two predictions and observed holdout data is computed, as it is for every other location, and these RMSE values are averaged across space and displayed over time in figure \ref{fig:S8}.

\end{document}